\pdfoutput=1
\documentclass[useAMS,usenatbib]{mn2e}
\usepackage{graphicx}
\usepackage{rotating}
\usepackage{lscape}

\title[The STREGA survey]{STREGA: STRucture and Evolution of the
  GAlaxy. I. Survey Overview and First Results\thanks{Based on data
    collected with the ESO INAF - VLT Survey Telescope and OmegaCAM at
    the European Southern Observatory, Chile (ESO Programme
    088.D-4015, 089.D-0706, 091.D-0623)}} 
\author[Marconi et al.]{M. Marconi$^{1}$\thanks{E-mail: marconi@na.astro.it},
  I. Musella$^{1}$, M. Di Criscienzo$^{1}$, M. Cignoni$^{2}$,
  M. Dall'Ora$^{1}$,  G. Bono$^{3}$, 
\newauthor
 V. Ripepi$^1$, E. Brocato$^4$, G. Raimondo$^5$,
  A. Grado$^1$, L. Limatola$^1$, G. Coppola$^1$, 
\newauthor 
M. I. Moretti$^1$, P. B. Stetson$^6$,
  A. Calamida$^{2,4}$, M. Cantiello$^5$,
  M. Capaccioli$^{7}$,
\newauthor
 E. Cappellaro$^8$,  M.-R.L. Cioni$^{9,10}$,
  S. Degl'Innocenti$^{11}$, D. De Martino$^{1}$, A.  Di
  Cecco$^{4,12}$, 
\newauthor 
I. Ferraro$^{4}$, G. Iannicola$^{4}$,
P. G. Prada Moroni$^{11}$,
  R. Silvotti$^{13}$, R. Buonanno$^{3,5}$,
\newauthor
 F. Getman$^{1}$,
  N. R. Napolitano$^1$, 
L. Pulone$^4$
and P. Schipani$^1$\\
$^{1}$INAF-Osservatorio Astronomico di Capodimonte, Salita
  Moiariello, 16, I-80131, Napoli, Italy\\
$^{2}$Space Telescope Science Institute, 3700 San Martin Drive, Baltimore, MD 21218, USA\\
$^{3}$Dipartimento di Fisica, Universit\`a degli Studi di Roma-Tor Vergata, via della Ricerca Scientifica 1, I-00133 Roma, Italy \\
$^{4}$INAF-Osservatorio Astronomico di Roma, Via Frascati 33, I-00044
Monte Porzio Catone, Italy\\
$^{5}$INAF-Osservatorio Astronomico Collurania, via M. Maggini, I-64100 Teramo, Italy\\
$^{6}$ NRC-Herzberg, Dominion Astrophysical Observatory, 5071 West Saanich Road, Victoria, BC V9E 2E7, Canada\\
$^{7}$Dipartimento di Fisica, Universit\`a ``Federico II'', Via
Cinthia, I-80126 Napoli, Italy\\
$^{8}$INAF-Osservatorio Astronomico di Padova, vicolo
dell'Osservatorio n. 5, 35122, Padova, Italy\\
$^{9}$University of Hertfordshire, Physics Astronomy and Mathematics,
College Lane, Hatfield AL10 9AB, UK\\
$^{10}$Leibnitz-Institut f\"ur Astrophysik Potsdam, An der Sternwarte 16, D-14482 Potsdam, Germany\\
$^{11}$Dipartimento di Fisica ``Enrico Fermi'', Universit\`a di Pisa, largo Pontecorvo 3, 56127, Pisa, Italy\\
$^{12}$Agenzia Spaziale Italiana Science Data Center (ASDC), c/o ESRIN, via G. Galilei, I-00044 Frascati, Italy\\
$^{13}$INAF - Osservatorio Astrofisico di Torino, via Osservatorio 20, 10025, Pino Torinese, Italy
}

\begin{document}

\date{Accepted ..... Received .... ;}


\maketitle

\label{firstpage}

\begin{abstract}
STREGA (STRucture and Evolution of the GAlaxy) is  a Guaranteed Time
 survey being performed at the VST (the ESO VLT Survey
Telescope) to map about 150 square degrees in the Galactic halo, in
order to constrain the mechanisms of galactic formation and evolution.
 The survey is built as a five-year project,
organized in two parts: a core program to explore the surrounding
regions of selected stellar systems and a second complementary part to
map the southern portion of the Fornax
orbit and extend the observations of the core program.
The adopted stellar tracers are mainly variable stars (RR~Lyraes and Long
Period Variables) and Main Sequence Turn-off stars
for which observations in the $g$,$r$,$i$  bands are obtained. 
We present an overview of the survey and some preliminary
results for three observing runs that have been completed. For
  the region centered on $\omega$~Cen (37 deg$^2$), covering about three tidal
  radii, we also discuss the detected stellar density
radial profile and angular distribution, leading to the identification
of  extratidal cluster stars.
We also conclude that the cluster tidal radius is about 1.2 deg, in
agreement with values in the literature based on the Wilson model.

\end{abstract}


\section{Introduction}

The formation and evolution of galaxies and their relation with the environment, from the scale of the Local Group to superclusters, has been a major field of astronomical research in the past decade.
The first laboratory for studying mechanisms of galactic formation and evolution
 and their dependence on the surrounding
environment is represented by the Milky Way (MW) and its satellite
galaxies. For nearby systems, indeed, accurate photometry can be obtained for
individual members of the various stellar populations. 
Several investigations have shown that the outer regions of
the Galactic halo do not have a smooth distribution of stars but
appear quite ``clumpy'' \citep[see e.g.][]{vz03,n03,bell10,drake13,zinn14}, in
agreement with theoretical models of the hierarchical formation of
structures based on a Cold Dark Matter cosmology
\citep[][]{navarro97,mateo98,dacosta99,benson02,tumlinson10,wang13}.  Among the most spectacular examples of
Galactic halo substructures we cite:
\begin{itemize}
\item The Sagittarius dwarf spheroidal
galaxy currently merging with the MW halo and its associated stream
\citep[see e.g.][and references therein]{Ibata94,Bellaz06,deg13};
\item the stellar
over-density in the Canis Major region whose origin is still a matter
of debate
\citep[][and references therein]{Martin04,Momany06,mateu09,lopez07,sav11}; 
\item the presence
of peculiar Galactic Globular Clusters (GGCs) with observed tidal tails or
suspected halos \citep[see e.g.][and references
therein]{fell07,oden01,oden09,olsze09,chun10,walker11,lane12a,lane12b,Maj12};
\item  the
discovery of several ultra-faint companions of the MW from 
analysis of the Sloan Digital Sky Survey (SDSS)  data \citep[][and
references therein]{Belokurov06,Belokurov07,Musella09,Musella12, Moretti09,
Clementini12,DallOra06,DallOra12,garofalo13}. Note that similar
substructures have been also found in the M31 halo
\citep[e.g.][]{gilb09,rich08,Clementini11}.  
\end{itemize}

Dating back to
\citet{Lynden76,Lynden82,DemKun79,Lynden95}, it was suggested that the
dwarf spheroidal satellites (dSphs) of the Galaxy and a number of its
globular clusters (GCs) are distributed along planar
alignments reflecting distinct orbital planes and usually interpreted
as the result of the disruption of larger galaxies.  The location of
these systems has recently been claimed to define a vast polar
structure \citep[VPOS;][]{paw12,paw13} and a similar plane has been
detected with very high significance in M31 \citep{conn13,ibata13}.  A
very interesting example of these alignments is the so called Fornax
stream that should include both dSphs (Fornax, Leo~I, Leo~II,
Sculptor and possibly Sextans and Phoenix ) and GCs (Pal~3,
Pal~4 and Pal~12), as also discussed in \citet[][]{Maj94}.
Even if the measurements of the absolute proper motion of Fornax
\citep[][]{Din04}, together with the shell structures identified by
\citet[][]{Col04}, already provided some circumstantial evidence,
accurate observations of large fractions of the Fornax orbit are still
lacking.  According to recent N-body simulations \citep[see e.g.][and
references therein]{Ma02,Ha03,Re06,Madau08,Ma11}, in the interaction
of small satellites (e.g. dSphs and GCs) with the MW, tidally stripped
material should form elongated tidal tails and spherical shells
extending many scale radii beyond the tidal radius.

 Evidence of tidal
tails has been found in connection with several GCs, for example Pal~14 
and NGC1851 \citep[see e.g.][and references therein]{sollima11,sollima12}, 
M15, M30, M53, NGC5053, NGC5466 \citep[][]{chun10} Pal~5
\citep[][]{Oden01} and also NGC5139 \citep[see e.g.][$\omega$~Cen]{Leon00,
 Maj12}.
This last cluster is of particular interest. It has been claimed to be the remnant of a tidally
disrupted galaxy because of: i) its internal chemical and age
distribution with the presence of multiple stellar populations;
ii) its  similarity
for mass and chemistry to M54, the GC associated with the core of the disrupting
Sagittarius galaxy \citep[see discussion in][]{Maj12} and iii) its unusual
low-inclination, retrograde orbit. 
Even if simulations of the tidal disruption of $\omega$~Cen suggested
the presence of stripped cluster stars in the Solar Neighborhood, the
observation of these extratidal stars only produced preliminary
results \citep[see e.g.][]{Leon00} that were later found to be
biased by substantial foreground differential reddening problems
\citep[][]{law03}.
Subsequently \citet{dacoco08} estimated that only a very small fraction
of  cluster members in the Red Giant Branch
(RGB) phase were actually stripped stars. \citet{Maj12} has recently reported
the detection of ``tidal debris'' consistent with that modeled
for $\omega$~Cen.

Another interesting GC hosting multiple stellar populations is NGC6752
\citep[see e.g.][and references therein]{carretta13,Milone2013}.
Even if the proper motion of this cluster has been extensively
investigated
\citep[see e.g.][]{dinescu99,drukier03,zlo12}, a
systematic investigation of extra-tidal stellar populations around this
cluster is lacking in the literature. 

The outer-halo sparse GC Pal~12 is probably younger
than the bulk of GGCs \citep[see e.g.][]{rosenberg98} and has
been suggested to have been tidally captured from the Sgr dSph galaxy
by the Milky Way \citep[][]{irwin99}. 
A tidal stream of debris from the Sgr dSph galaxy towards the Galactic halo
has been observed both by the Two Micron All Sky Survey \citep[2MASS][]{maj03}
and by the SDSS \citep{ivezic04} and various authors have shown that Pal~12 is embedded in
this stream \citep[see e.g.][]{Bellaz03,md02,cohen04}.

Finally, it is worth noting that signatures of extratidal stellar populations have been detected around a
number of nearby dSphs \citep[e.g.][and references
therein]{ih95,Maj00,mateo08,paw12}. A typical example is the Carina
galaxy for which several authors have investigated the presence of
tidal tails \citep{kuhn96,monelli04,munoz06}.

In this context, the STRucture and Evolution of the GAlaxy (STREGA)
survey plans to use the VLT Survey Telescope (VST) to investigate Galactic halo formation
mechanisms through the two following observing strategies: i) tracing tidal
tails and halos around stellar clusters and galaxies; ii) mapping
extended regions of the southern portion of the Fornax orbit to
trace the Fornax Stream. In addition, our
strategy can in principle allow us to identify unknown very faint
stellar systems by using well established techniques developed in the
context of the SDSS \citep[e.g.][]{Belokurov06}.

In this paper, we describe the instrument, namely the VST and OmegaGAM
(Section \ref{VST}) and the STREGA
survey (Section \ref{survey}).  In  Section \ref{sec-data}, we
present the observations for the three completed runs, covering the
GGCs $\omega$~Cen,
NGC6752 and Pal~12, whereas the procedures used for the data reductions are
discussed in Section \ref{reduction}. The adopted tools and some
preliminary results for the fields around NGC6752 and $\omega$~Cen  are shown in Section
\ref{sec-results}. The Conclusions and some final remarks close the paper.

\section{Observing with VST}\label{VST}

The VST is a 2.6-m wide-field optical telescope built
by INAF-Osservatorio Astronomico di Capodimonte, Naples (Italy) 
and located at Cerro Paranal (Chile) on the VLT
platform. It is characterized by an alt-azimuth mount with a
f/5.5 modified Ritchey-Chr\'e tien optics \citep[see][for further details]{VST}.
The telescope is equipped with OmegaCAM \citep{ku11}, a camera with  a field of view of $1^{\circ}$ x $1^{\circ}$, built by a consortium
of European institutes, featuring a 32-CCD, 16k x
16k detector mosaic, with a pixel size of 0.21 arcsec.
The filter system includes the USNO $u'g'r'i'z'$
bands, the Johnson $B$ and $V$
filters, and a number of 
narrow-band filters mosaics such as segmented $H\alpha$ and $vS$. 
It is important to note that even if the VST filter system is based
on the USNO $u'g'r'i'z'$ system \citep{smith02}, slightly
different from SDSS $ugriz$ (for details, see the “Photometry White
Paper” by Gunn et al. 2001), we calibrate our data to the ``natural'' SDSS
system. For this reason, in the following, we use the $ugriz$
notation. 
Finally, we remind the reader that the OmegaCAM calibration plan ensures for the
VST  data a photometric and astrometric calibration at 0.05 mag and
0.1 arcsec rms precision levels, respectively.

\section{The survey STREGA@VST} \label{survey}

\begin{figure*}
\includegraphics[width=0.9\textwidth]{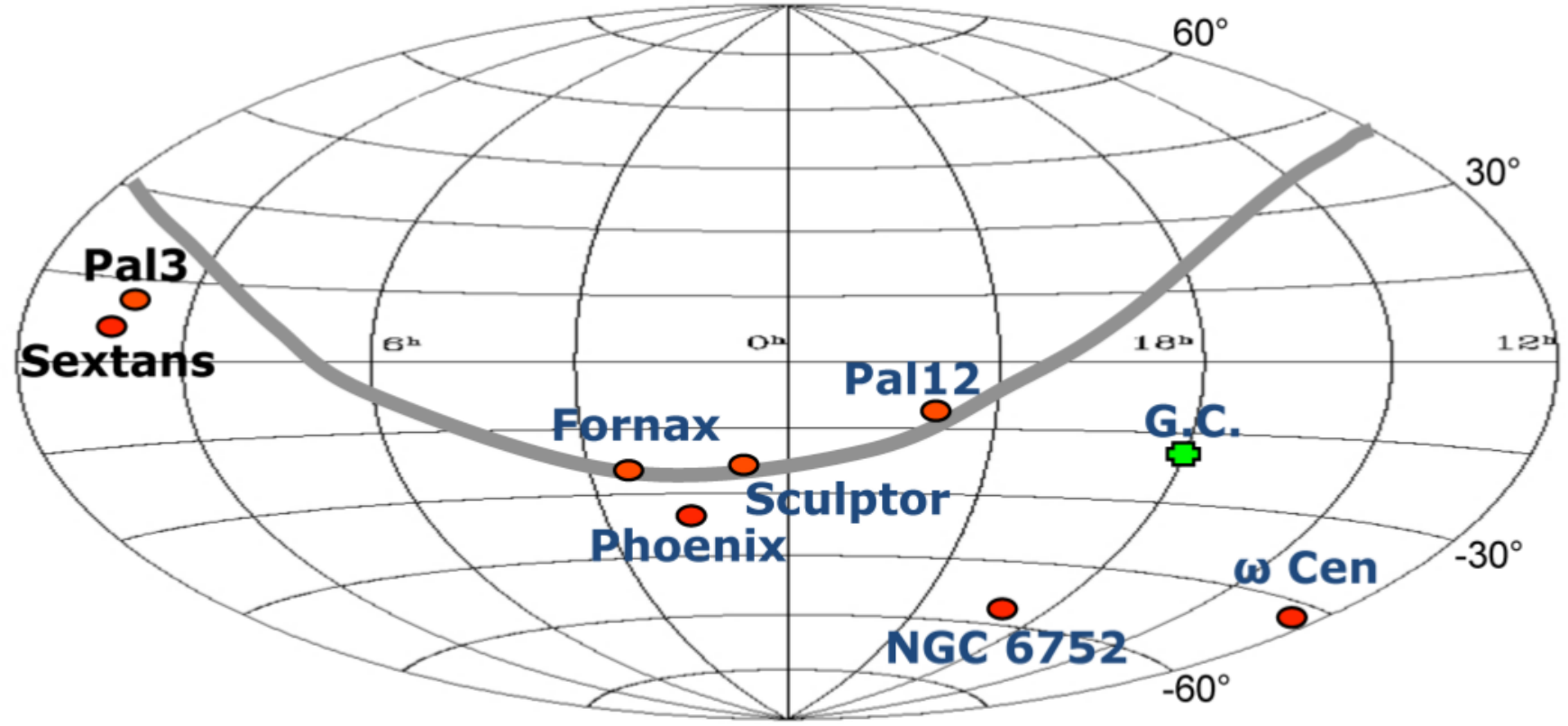}
\caption{The selected STREGA core program targets (red symbols) in
  equatorial coordinates, compared with the location of the Galactic
  Center (green symbol) and Fornax orbit according to \citet{Din04}.} \label{map}
\end{figure*}

STREGA@VST \citep[P.I.: M. Marconi/I. Musella, see also][]{m14} was conceived as a five-year survey,
organized in two parts: a core program and a second part. In the
core program we are exploring the outskirts of selected dSphs
and GCs up to at least
2--3 tidal radii in several directions to distinguish between tidal tails
and halos. These systems are  located either along the Fornax orbit or are
of particular interest for the interaction mechanism with the Galactic
halo: Fornax and Sculptor (38 fields), Phoenix (3 fields), Sextans (13
fields), Pal~3 (3 fields), Pal~12 (1 field), $\omega$~Cen and NGC6752
(33 and 32 fields around each GC, respectively). 
In Fig.~\ref{map} we show the location of the core program targets
(red symbols).


 The second part will
cover strips of adjacent fields distributed transverse to Fornax
orbit and will extend the observations of the most interesting systems
explored in the core program to larger radial distances.

To investigate the mechanisms of the formation and evolution of the
Galactic halo, the STREGA survey mainly relies on variable stars (RR~Lyrae 
stars and Long Period Variables) and Main Sequence Turn-off (MSTO) stars \citep[see
e.g.][]{Cignoni07}. The former are easily detectable thanks
to their brightness and characteristic light curves \citep[see
e.g.][]{VZ06,mateu09,Prior09}. The latter are about 3.5 mag fainter than RR~Lyrae
stars but at least 100 times more abundant.
Moreover, the
accurate sampling of the light curves of the identified RR~Lyrae stars is used to characterize their stellar properties providing
clues on the Galactic halo star formation history \citep[][and
references therein]{Clementini11}.  The multi-purpose nature of the
STREGA survey aims at a few additional objectives including the
sampling of the 
brightest field single
white dwarfs (WDs) and interacting binaries (IBs), most being
accreting compact objects with low-mass donor companions. 
In this respect our survey complements the VST Photometric H$\alpha$
Survey of the Southern Galactic Plane \citep[VPHAS+][]{drew14} that will map the
disk population of IBs. 
As for the single WDs, our relatively deep fields at different
Galactic latitudes will allow us to study the poorly known WD parameter space
in the various Galactic populations, including thin disk, thick disk and
spheroid populations, complementing the recent statistical studies
based on much larger, but shallower, WD samples from the SDSS and
SuperCOSMOS surveys \citep[][]{dege08,rh11}.  We will use different
filter strategies, with the narrow $vS$ filter to sample hot WDs and
the broad SDSS filters
for cool/ultracool WDs. Similar to the case with IBs and especially for the halo
candidates, multi-epoch observations and follow-up IR photometry
and/or spectroscopy will allow us to confirm the true nature of these
objects.

\section{The Observations}\label{sec-data}

Observations for the VST INAF GTO started at the end of 2011 and
proposals have been approved for Periods from 88 to 93 (088.D-4015,
089.D-0706, 090.D-0168, 091.D-0623, 092.D-0732, 093.D-0170). However,  also due to technical and scheduling problems,
only a minor part of the observations for the STREGA core program have been gathered so far (see Table
\ref{grade_oss}).  In particular the only  completed patches are the
fields around $\omega$~Cen (observed in the ESO Period 88), NGC6752
(Period 89) and Pal~12 (Period 91). Additional data for the central region of
$\omega$~Cen (4 fields) were
observed subsequently to P88 in visitor mode (March 2013).

\begin{table}
\caption{Distribution of the service mode observations. The first
  column indicates the ESO Period. The second and
third columns give the requested and observed time, respectively. The
last column reports the ESO OB class percentage.} \label{grade_oss}
\label{esoPeriods}
\begin{center}
\begin{tabular}{cccc}
\hline
\hline
\noalign{\smallskip} 
ESO Period & Requested [h] & Observed [h]& \% OB Class A,B,D  \\
\noalign{\smallskip}
\hline
\noalign{\smallskip} 
88  &   28 & 3.8  & 26,67,7  \\
89  &  38.4 & 3.8 & 22,76,2 \\
90 &   47.9 & 9.4 & 50,39,11 \\
91 &  39.6 & 8.9 & 38,37,25 \\
92 &   18.0  & 11.7 & 59,37,4   \\
93 &  33.0 & OPEN & ---\\
\hline
\end{tabular}
\end{center}
\end{table}

\begin{figure*}
\includegraphics[width=0.9\columnwidth]{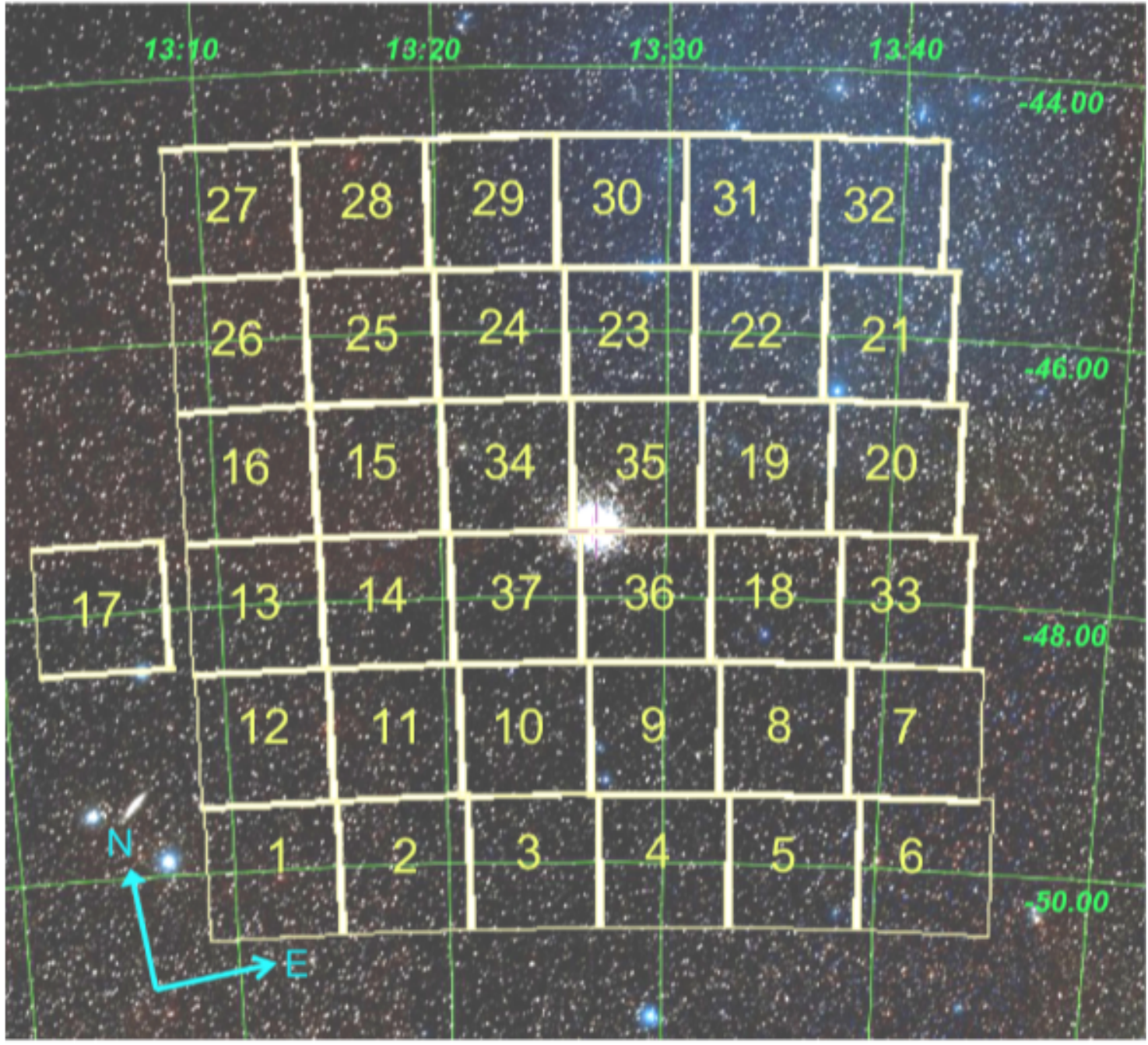}
\caption{Observed fields by OMEGACAM@VST around the Globular Cluster
  $\omega$~Cen, in equatorial coordinates.} \label{f:omegacen}
\end{figure*}

\begin{figure*}
\includegraphics[width=0.9\columnwidth]{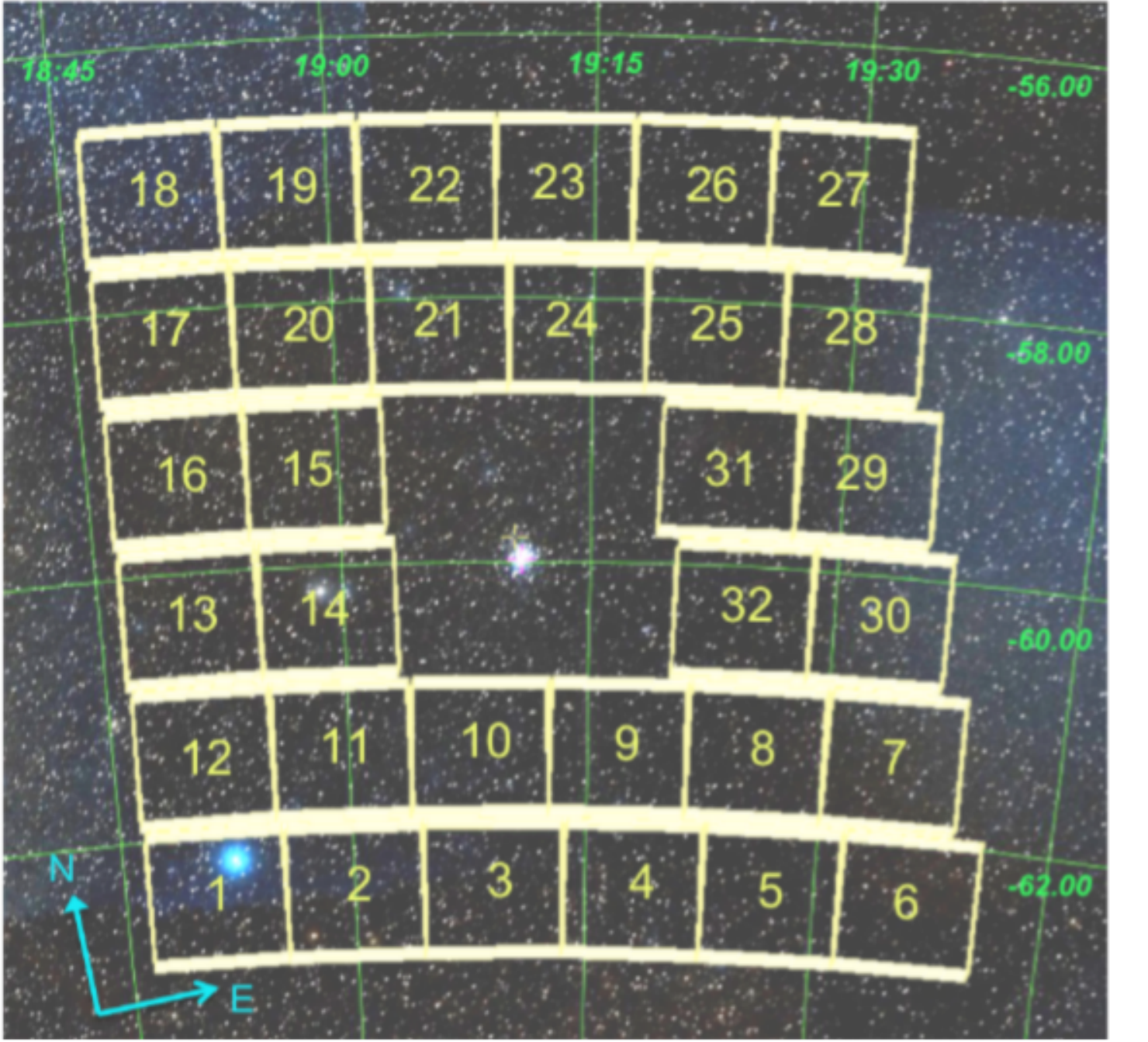}
\caption{As in Fig.~\ref{f:omegacen} but for NGC6752.} \label{f:ngc6752}
\end{figure*}

Due to the relatively short distance modulus of the globular clusters
$\omega$~Cen \citep[$\approx$13.7 mag;][]{DelPrincipe2006}, NGC6752 
\citep[$\approx$13.2 mag;][]{Milone2013} and Pal~12 
\citep[$\approx$16.5 mag;][]{rosenberg98}, the exposure times to reach the 
RR~Lyrae magnitude level
in these systems are of the order of a very few seconds, dramatically shorter
than telescope overheads. In these
cases variability should be investigated with suitable follow-up observations on the most interesting
fields. On the other hand, thanks to its wide field of view and high
resolution, VST is the ideal instrument to perform a
map of extra-tidal stellar populations or of the extended halo around
these systems, through observations down to the MSTO and fainter MS
stars. Therefore, we planned to observe the selected fields to deeper
magnitude limits (exposure times of tens of seconds) to build color-magnitude and color-color diagrams.  The
fields observed for  $\omega$~Cen and NGC6752 were selected to
uniformly map the outskirts of the cluster up to 3 tidal
radii, by adopting the Survey Area Definition Tool
\citep[SADT;][]{Arnaboldi08}. In Figs. \ref{f:omegacen} and
\ref{f:ngc6752}, we show the obtained
sky field distributions. For Pal~12, we observed a single field centered
on the cluster, covering about 2 tidal radii.\\


\section{Data reduction}\label{reduction}

The  data reduction  has made  use of  the newly  developed VST--Tube
imaging  pipeline \citep{grado}, specifically  conceived for  the data
coming from the VST  telescope but adaptable to other
existing or future single or multi-CCD cameras.  The data processing includes
removal of instrumental signatures, namely overscan, bias and flatfleld
correction, CCDs gain harmonization and illumination
correction. Relative and absolute astrometric and photometric
calibration were applied to the individual exposures.
In detail, the bias level is measured from the median value of a
suitable portion of the overscan region, then the bias map (masterbias) is derived
after a
sigma-clipped averaging of the bias frames. The flat-field correction (masterflat)
is a combination of the average twilight flats (from a number
of twilight flats), that correct for pixel-to-pixel sensitivity
variation,  and a super skyflat, obtained with a combination of
science images, that accounts for the low spatial frequency gain
variations.  In order to set all the mosaic
chips to the same zeropoint, a gain harmonization procedure was required. The procedure
measures the relative gain coefficients that give the same background
signal in adjacent CCDs.  For the VST data it was necessary to apply
an additional correction for the scattered light. This is a common
problem for  wide-field imagers where telescope and instrument  baffling can
be an issue. The off-axis uncontrolled  redistribution  of  light paths
gives an  additive component to the background, with a peak in the
central area, so that the flat field
will  not be an  accurate estimate of the spatial detector response.
Indeed, if this effect is not corrected, the image background  will appear
perfectly flat,  but the  photometric response will be  position
dependent \citep{IC}.  This error in  the determination  of the
flat-field can be mitigated through the determination and application
of the so called illumination correction (IC) map. The IC map was determined by comparing the magnitudes of equatorial photometric standard fields
stars with extracted catalogs from the SDSS DR8.  The magnitude residuals
as a function of the position were fitted using a generalized adaptive
method (GAM) in order to obtain a 2D map used to correct the
science images during the pre-reduction stage. The GAM allows us to
obtain a good surface also in case the field of view is not
uniformly sampled by the standard stars.  The absolute photometric
calibration (see Table \ref{calibLog}) was computed comparing the aperture
magnitude \citep{sex} of photometric standard-field stars  with
reference catalogs from SDSS DR8. A simultaneous fit of the zero-point
and color term was performed using Photcal \citep{radovich04}, while the
extinction coefficients were those (mean values) extracted from the extinction curve
provided by the ESO calibration team.   Relative photometric
corrections, accounting for transparency variations among exposures,
 as
well as  absolute and relative astrometric calibration, were obtained
using SCAMP \citep{scamp}.  Resampling and stacking of the
individual exposures was obtained  using SWARP
 \citep{swarp}.


   \begin{table*}
   \begin{center}
      \caption[]{Calibration Log. The table reports the observed
        target, the adopted filters, the night used for the absolute
        photometric calibration and the corresponding zero point, color term
        and extinction coefficient}
         \label{calibLog}
         \begin{tabular}{c c c c c c }
            \hline\\
            \noalign{\smallskip}
            Target   & Filter & night &  Zero-Point (mag) & Color term
            & Ext. coeff. (mag)  \\
            \noalign{\smallskip}
            \hline\\
            \noalign{\smallskip}
            $\omega$~Cen & g &  2012-03-21 & 24.650 $\pm$ 0.010
            & 0.024 $\pm$ 0.009 $(g-i)$ & 0.110 \\
            $\omega$~Cen & r &  2012-04-02 & 24.502 $\pm$ 0.008
            & 0.043 $\pm$ 0.020 $(r-i)$ & 0.030\\
            $\omega$~Cen & i &  2012-03-04 & 24.013 $\pm$ 0.011
            & -0.001 $\pm$ 0.008 $(g-i)$ & 0.010 \\
            $\omega$~Cen & g &  2013-03-15/19 & 24.699 $\pm$ 0.006 &  0.024 $\pm$ 0.006 $(g-i)$& 0.180 \\
            $\omega$~Cen & r &  2013-03-15/19 & 24.601 $\pm$ 0.005 &  0.049 $\pm$ 0.013 $(r-i)$& 0.100  \\
            $\omega$~Cen & i &  2013-03-15/19 & 24.119 $\pm$ 0.006 &  -0.002 $\pm$ 0.005 $(g-i)$& 0.043 \\
            NGC6752 & g &  2012-05-05 & 24.799 $\pm$ 0.010 &
            0.017 $\pm$ 0.010 $(g-i)$ & 0.180  \\
            NGC6752 & r &  2012-05-01 & 24.598 $\pm$ 0.006 & 0.054 $\pm$ 0.016 $(r-i)$& 0.100  \\
            NGC6752 & i &  2012-05-08 & 24.136 $\pm$ 0.006 & -0.002 $\pm$ 0.016 $(g-i)$& 0.043 \\
            PAL 12 & g &  2013-07-01 & 24.809 $\pm$ 0.008 &
            0.033 $\pm$ 0.007 $(g-i)$ & 0.180  \\
            PAL 12 & r &  2013-07-01 & 24.633 $\pm$ 0.006 &
            0.053 $\pm$ 0.017 $(r-i)$ & 0.100 \\
            PAL 12 & i &  2013-07-01 & 24.113$\pm$ 0.007 &
            -0.002 $\pm$ 0.006 $(g-i)$ & 0.043 \\
            \noalign{\smallskip}
            \hline\\
	\end{tabular}
	\end{center}
   
   \end{table*}

The stellar photometry in the crowded stellar clusters must be performed
by using the Point Spread Function (PSF) fitting method
\citep[e.g.][]{stetson1987}, while for uncrowded fields, to obtain accurate magnitudes, we could also use
aperture photometry. In practice, for crowded
fields we use \texttt{DAOPHOT/ALLSTAR} \citep{stetson1987}, one of the
most used packages to obtain very accurate PSF magnitudes, and for
measuring aperture photometry we adopt \texttt{SExtractor} \citep{sex}. The
latter is currently adopted for extragalactic studies, but it is
particularly suited for wide-field images, being fast, fully
automatisable and able to give accurate results for stellar
photometry in not severely crowded fields.  For each source,
\texttt{SExtractor} provides several different parameters to
characterize the brightness and the shape.  The proper aperture
magnitude for stellar objects is the so called 
\texttt{MAG\_APER} parameter. Indeed it gives a measure of the total magnitude within
a circular aperture with a fixed user-supplied radius. In our analysis, we
adopt an aperture radius of 20 pixels (correposnding to $\sim 5 \times
FWHM$ for typical seeing) that
encloses more than 99.99$\%$ of the light of a selected, isolated and
good number of stars among the brightest ones.  Moreover, to
check the consistency of the  photometry we compared  the
\texttt{SExtractor} with the \texttt{DAOPHOT/ALLSTAR}
 measurements, for selected images.

To clean the catalogs obtained by \texttt{DAOPHOT/ALLSTAR} of
objects with noisy photometry, we
selected the objects with $\chi<1.6$ and
$-0.2<$\texttt{Sharp}$<0.2$
\footnote{In the single-frame photometry file, the
  \texttt{DAOPHOT/ALLSTAR} $\chi$ parameter of
each star is a robust estimate of the observed pixel-to-pixel scatter
of the fitting residuals to the expected scatter, whereas the
\texttt{Sharp} parameter is related to the intrinsic angular size of
the astronomical objects, and for stellar objects should have a value
close to zero.}.

\begin{figure*}
 \includegraphics[width=0.9\textwidth]{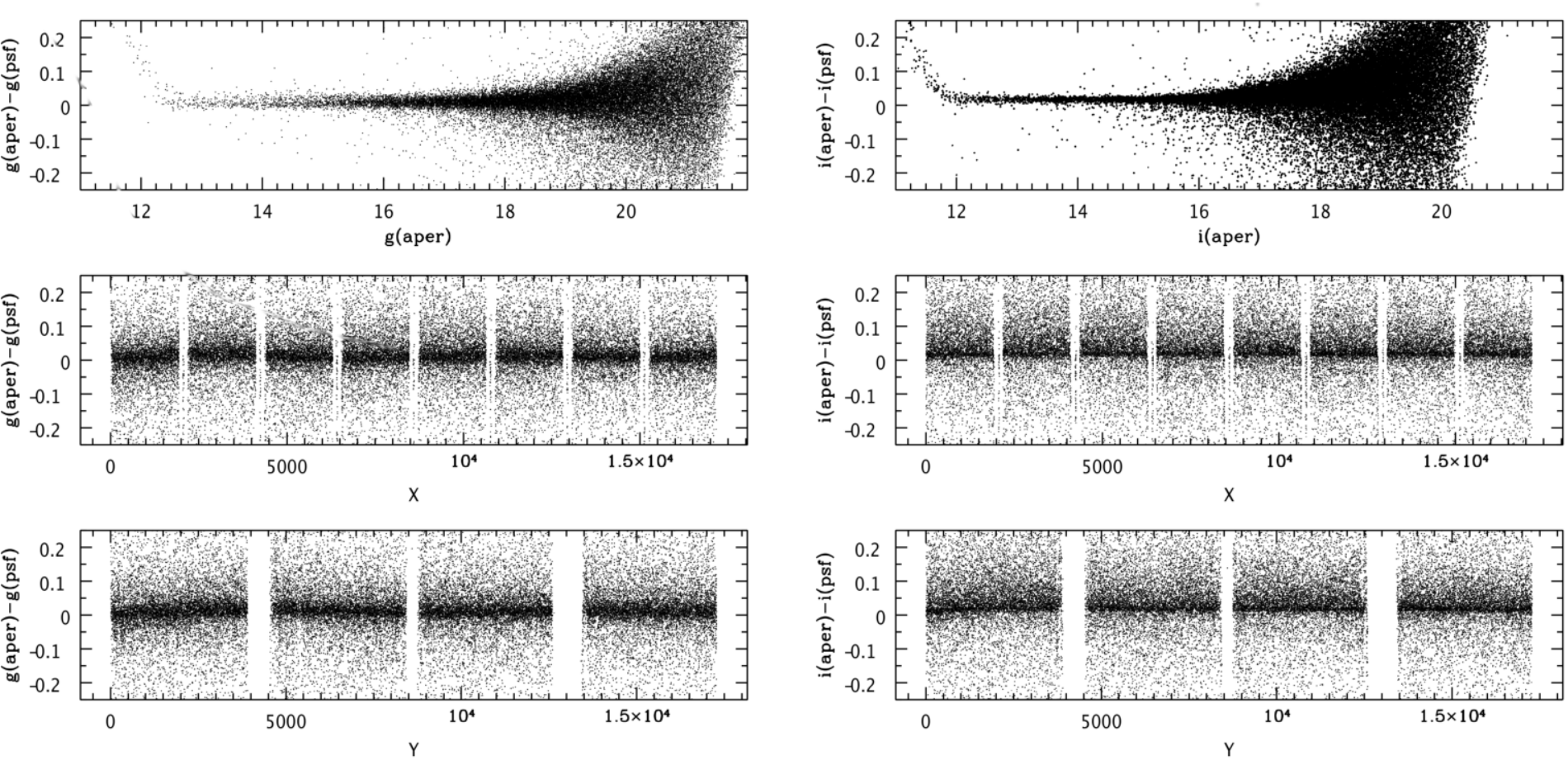}
 \caption{Residual between $g$ (left panel) and $i$ (right panel) magnitudes estimated by \texttt{SExtractor}
   and the PSF ones obtained with \texttt{DAOPHOT/ALLSTAR}
   (accounting for the different instrumental magnitude zero-point)  as a function  of the aperture
   magnitude and of the position on the frame, in the case of  the
   selected reference field ($l$,$b=$306.2,12.78; Field 1 around
   $\omega$~Cen).}
 \label{SEXvsDAOPHOT}
\end{figure*}

Fig.~\ref{SEXvsDAOPHOT} shows the residuals of the \texttt{SExtractor}
and \texttt{DAOPHOT} photometry, as a function of the magnitude and
the field position in the case of a  selected pointing ($l$,$b=$306.2,12.78; Field
\#1 around $\omega$~Cen), both in $g$ (left panel) and $i$ (right panel)
bands. We note that \texttt{SExtractor} aperture
photometry, in the case of uncrowded fields, is comparable with the more time-consuming \texttt{DAOPHOT} PSF
photometry, at least in the magnitude range 12.5--21 mag, with a maximum
discrepancy of 0.2 mag. At brighter magnitudes the deviation, more
evident in the $i$ band, is due to nonlinearity of the CCDs. 

Through this comparison between the aperture \texttt{SExtractor}
photometry and that obtained by \texttt{DAOPHOT/ALLSTAR} we also verified that,  in the case of the \texttt{SExtractor} catalog, the
contamination by non-stellar objects affects a magnitude range fainter
than the completeness limit (see Sect. \ref{preliminary}) and
therefore out of interest for our purposes.

The photometry on the fields  including the central part of
$\omega$~Cen  (\#34, \#35, \#36 and  \#37) and on the field centered on Pal~12 
was performed only using the \texttt{DAOPHOT/ALLSTAR} packages. The
calibration of the four central pointings on $\omega$~Cen, observed on a
non-photometric night, was checked with the deep and accurate $UBVRI$
photometry published in \citet{Castellani07}, transformed to the
$ugriz$ photometric system by adopting the equations computed for the
Population II stars, based on the transformations devised by
\citet{Jordi06}\footnote{http://www.sdss.org/dr7/algorithms/sdssUBVRITransform.html}. For
each field and for each filter, the mean offset between our measured
magnitudes and the transformed magnitudes was applied as a zero-point
correction. The four corrected catalogs were merged into a single
catalog, which is used for the subsequent analysis. Similarly for the
field centered on Pal~12, we checked the photometric calibration by
comparison with independent very accurate $BVI$ photometry by one of
us (P.B. Stetson, hereinafter PBS, unpublished). In the case of NGC6752, waiting for the VST
observations of the central fields, we performed a qualitative
comparison of the color-magnitude diagrams for the external fields againts the central one,
the latter obtained from
unpublished photometry by PBS, finding a satisfactory agreement
(see the following section).

\section{Color-Magnitude Diagrams}\label{sec-results}


\begin{figure*}
\centering
\includegraphics[width=0.9\textwidth]{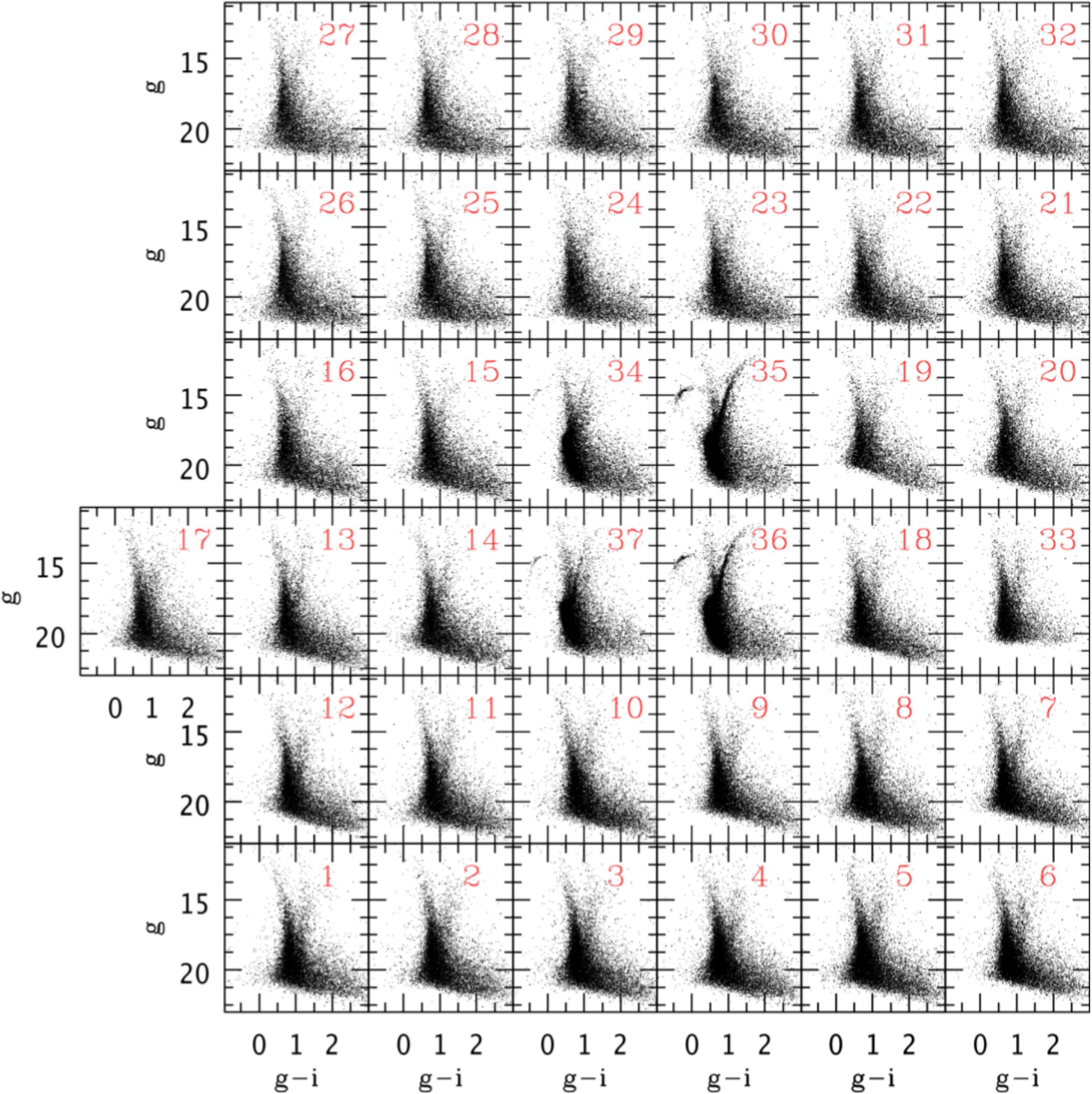}
\caption{ CMDs of the stars in the 37 fields on and around
  $\omega$~Cen. For orientation of the fields, see Fig. \ref{f:omegacen}}
\label{CMD_OMEGACEN}
\end{figure*}


\begin{figure*}
\centering
\includegraphics[width=0.9\textwidth]{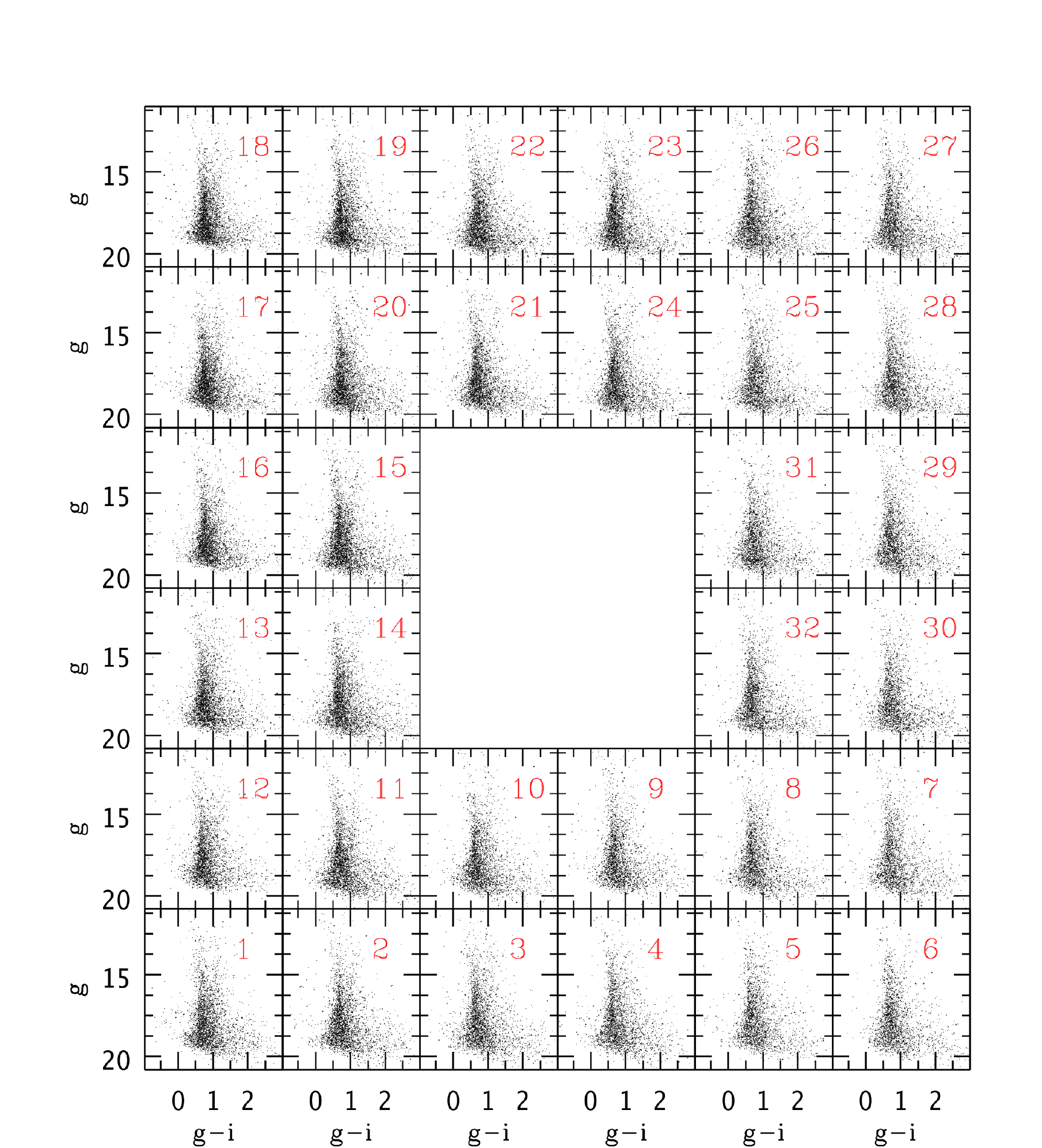}
\caption{ CMDs of the stars observed  in the 32 fields around NGC6752. For orientation of the fields, see Fig. \ref{f:ngc6752}.}
\label{CMD_NGC6752}
\end{figure*}


\begin{figure*}
\centering
\includegraphics[width=0.9\columnwidth]{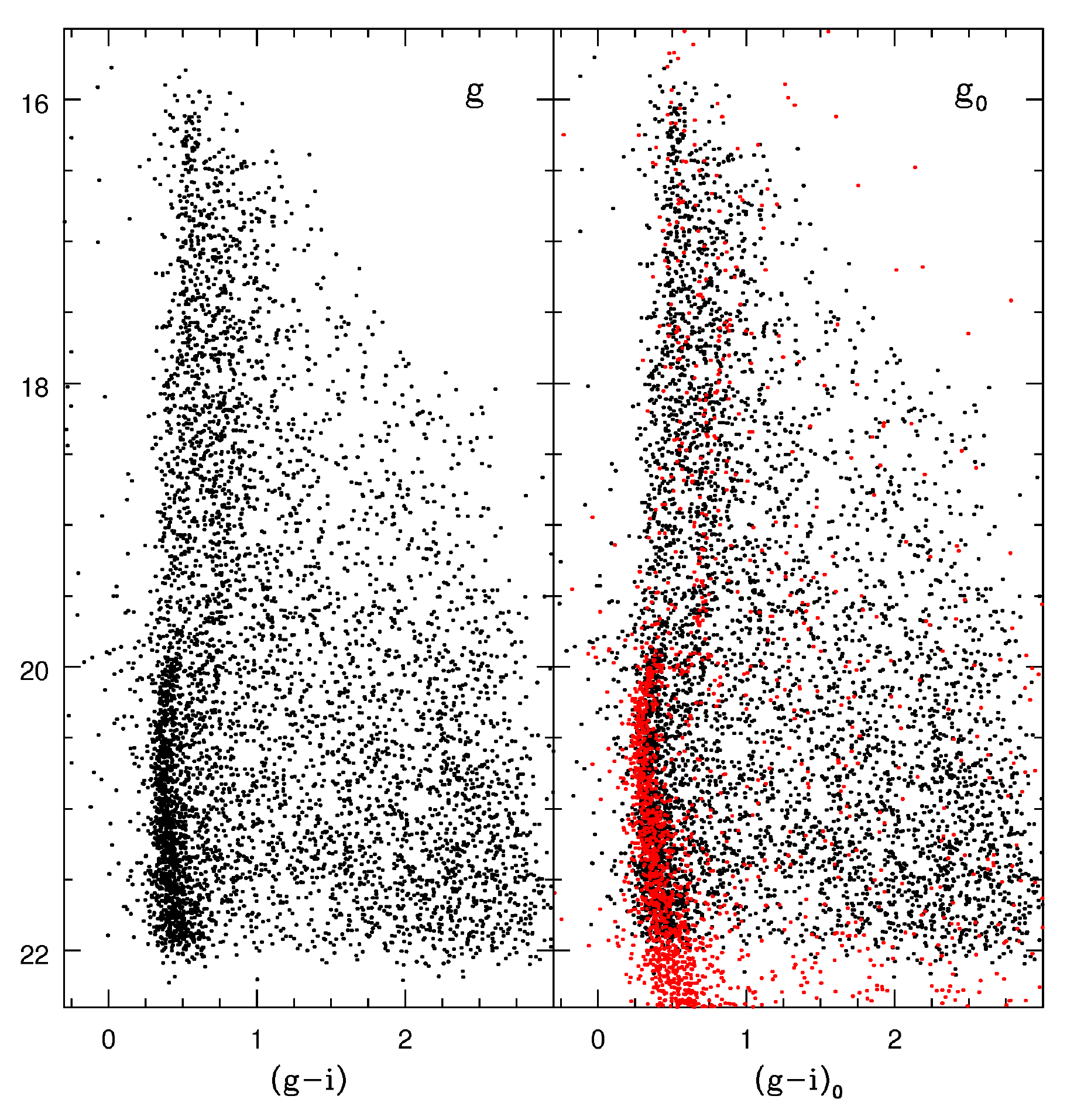}
\caption{CMD for the  field centered on Pal~12, before (left panel) and
  after (right panel) reddening correction. The reddening correction
  is negligible, but in the right panel, we
  also show the cluster photometry obtained by PBS (red
  symbols, see text
  for details).}
\label{CMD_PAL12}
\end{figure*}

Figures \ref{CMD_OMEGACEN}, \ref{CMD_NGC6752} and the left panel of
\ref{CMD_PAL12} show  the color-magnitude diagrams (CMDs) for $\omega$~Cen, 
NGC6752 and Pal~12, in all the investigated fields, obtained
by combining the  two independently selected $g$-band and $i$-band catalogs and
adopting a 0.25 arcsec matching radius. 

To investigate the signature of over-densities associated with various
CMD features, it is important to take into account the differential
reddening contribution as well as the contamination from the various
Galactic components.

\subsection{The differential reddening issue}


\begin{figure*}
\includegraphics[width=0.9\columnwidth]{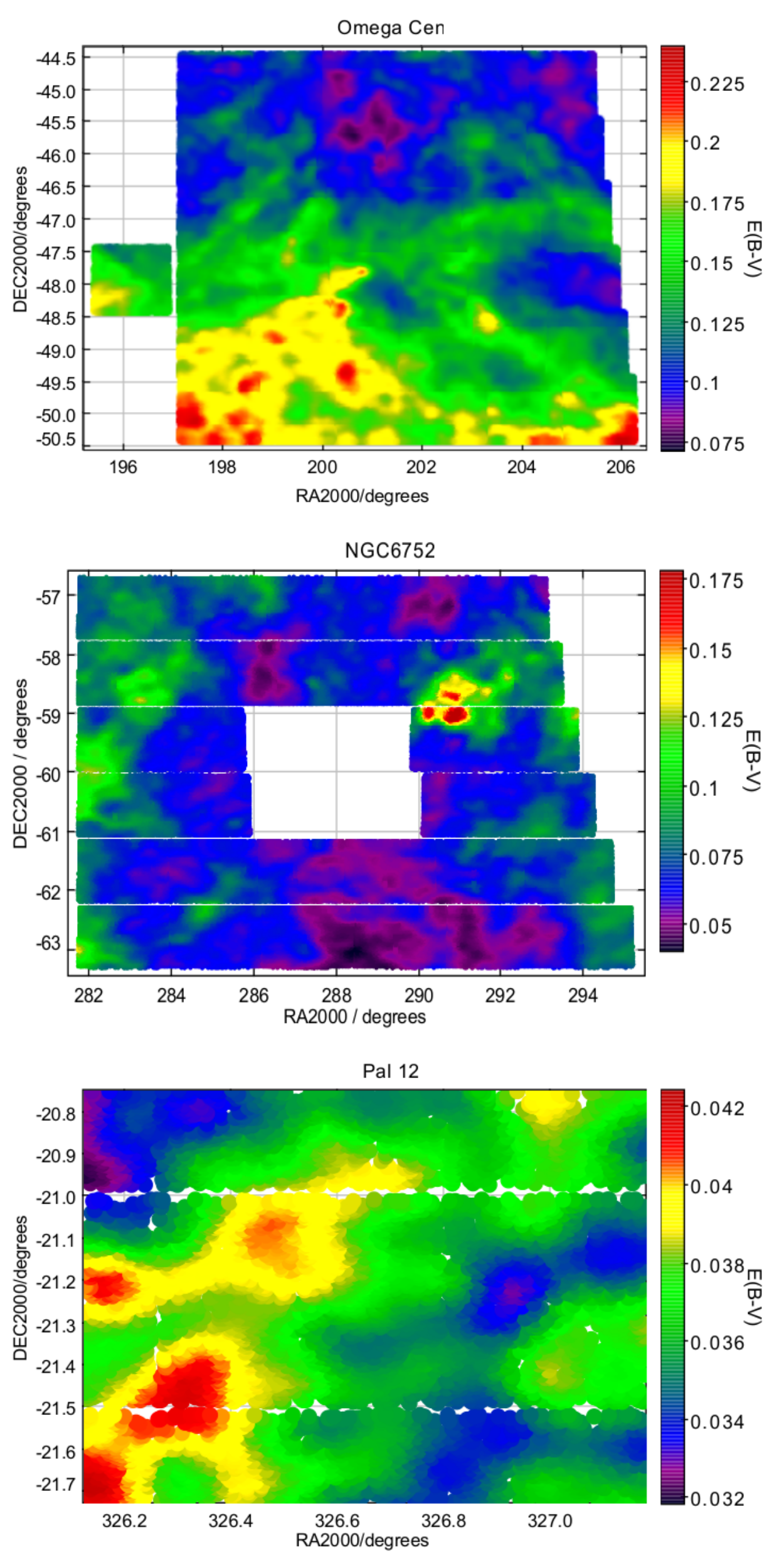}
\caption{The extinction maps of
\citet{schlegel1998} recalibrated by \citet{schlafly2011} for the observed
fields around $\omega$~Cen, NGC6752 and Pal~12.}
\label{reddening}
\end{figure*}

In Fig.~\ref{reddening} we report the extinction maps of
\citet{schlegel1998} recalibrated by \citet{schlafly2011} for the observed
fields. Significant differential reddening appears
only around $\omega$~Cen (note the differences among the color scales), due to the closeness of the lowest fields to
the Galactic disk. For NGC6752 and Pal~12, the reddening is rather
uniform apart from a small region around NGC6752 close to the
Galactic disk.


\begin{figure*}
\includegraphics[width=0.9\columnwidth]{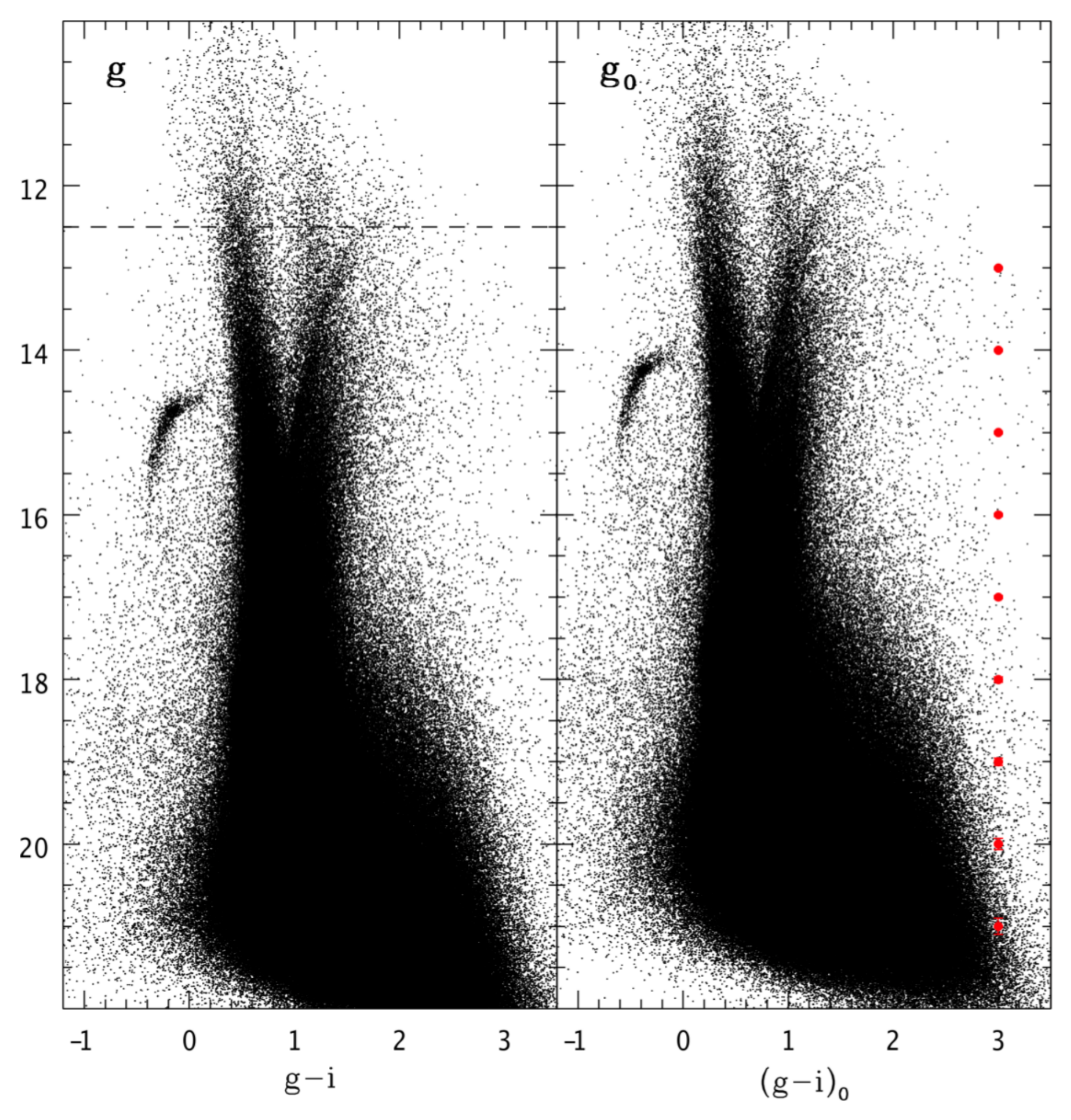}
\caption{Cumulative CMD for $\omega$~Cen before
  (left panel) and after (right panel) reddening correction. The red
  symbols represent the errors at the various luminosity levels. The
  dashed line represents the CCD nonlinearity  limit (see
  Fig.~\ref{SEXvsDAOPHOT} and text for details)}
\label{CMDtuttoOMEGA}
\end{figure*}


\begin{figure*}
\vspace{1truecm}
\includegraphics[width=0.9\columnwidth]{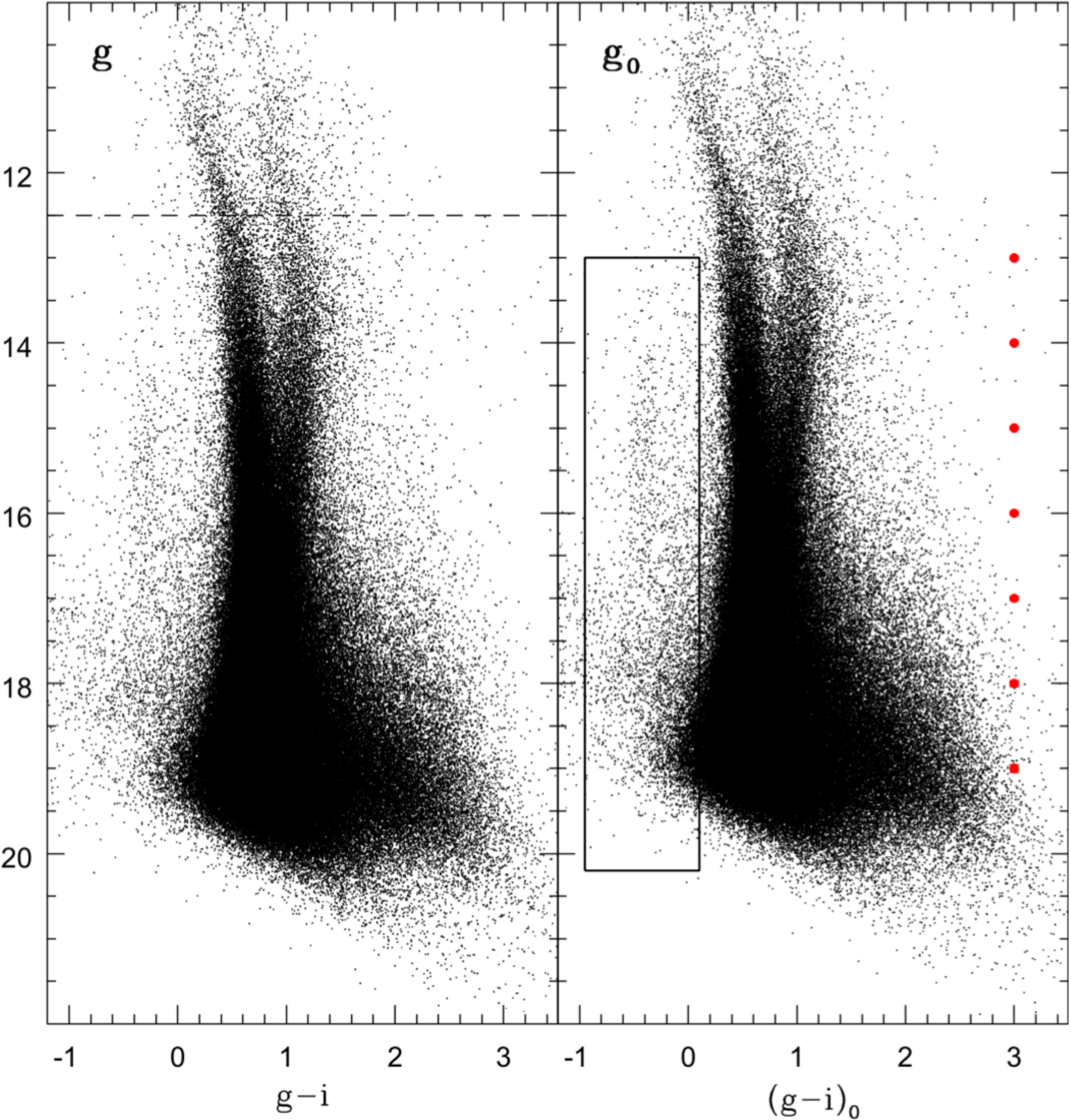}
\caption{As in Fig.~\ref{CMDtuttoOMEGA} but for NGC6752. The box
  encloses the vertical structure discussed in the text.}
\label{CMD_tuttoNGC6752}
\end{figure*}

We have used the Schlegel's maps  to obtain a merged total CMD for the
surroundings of  each cluster with de-reddened magnitude and
colors. In Figs.  \ref{CMDtuttoOMEGA} and \ref{CMD_tuttoNGC6752} we
report the cumulative CMDs before (left) and after (right) reddening
correction.  The dashed line indicates the limit of the CCD 
nonlinearity quoted above.

In the case of Pal~12 (see the bottom panel of Fig.~\ref{reddening})
the reddening differences are expected to be very small and we can
assume $\Delta$E(B-V) $\sim$ 0 over the entire field. In any case,
in the right panel of  Fig.~\ref{CMD_PAL12} we report the de-reddened
CMD (black dots) compared with the independent photometry by Stetson
(red dots; for details see above). The cluster MSTO is clearly
observed from this plot, but a detailed analysis is required in order
to disentangle the claimed contribution of the Sagittarius galaxy
CMD. For this reason,  the investigation of the stellar populations
within the observed two tidal radii for this cluster is presented in a
separate paper (Di Criscienzo et al. in prep).

For $\omega$~Cen and NGC6752, we checked the internal photometric
alignment by using the overlap region of adjacent frames and through
over-imposition of the Galactic component features (e.g. the Galactic
halo MSTO)  in the CMDs within about $\pm$ 0.05 mag.

\subsection{Galactic and synthetic cluster model predictions}

The observed fields in the direction of $\omega$~Cen and NGC~6752 have
low galactic latitude, so that the lines of sight are mainly
contaminated by Galactic thin and thick disk stars, with only a minor
contribution of halo stars. Although a quantitative
  interpretation of these populations is beyond the goal of this paper,
  the simple comparison of the data with the predictions of a Galactic
  model can be illuminating in the interpretation of the observed CMDs. To this aim we have
  used an updated version of the code described in
\citet{Castellani2002}, which assumes three galactic components,
namely a thin disk, a thick disk and a stellar halo, with specific
spatial structures and star formation laws. The code relies on the
latest stellar tracks by \citet{bressan2012} for normal stars and the
theoretical cooling tracks by \citet{salaris2000} for the white dwarf
phase. In this model the Galaxy is described by the following
ingredients:

\emph{Spatial structure:} the thin and thick disk are both described
as double exponentials: in Galactocentric distance in the Galactic
plane and in height over the plane. The respective scale heights are
250 pc and 750 pc, while the scale lengths are 3 kpc and 3.5 kpc. The
halo follows a power-law decay with exponent 3.5
\citep[][]{cabrera05,chen11}.  Thick disk and halo normalizations
  relative to the thin disk component are 12\% \citep{Juric2008} and 1/500 \citep[see
    e.g.][]{bahcall84,Tyson88}, respectively;

\emph{Star formation laws:} for each component the star-formation rate
is assumed constant. Stellar ages vary between 0 and 9 Gyr in the thin
disk, between 9 and 11 Gyr in the thick disk, and between 11 and 13
Gyr in the halo. The corresponding average metallicities are $Z=0.02$, $Z=0.006$ and
$Z=0.0002$\footnote{The adopted metallicity for the Galactic
    halo is a lower limit of current empirical estimates \citep[see
      e.g.][and references therein]{carollo07}. However, the goal here
    (see next Section) is not to generate a comprehensive halo model,
    but only to test the color visibility, in the CMD, of $\omega$ Cen and
    NGC6752 MSTOs with respect to the field. In fact, a more metal
    rich halo would have redder colors, thus enabling a still easier 
    detection of $\omega$ Cen and NGC6752 MSTOs.}, respectively;

\emph{Initial Mass Function:} all the galactic components assume a
Kroupa 2001 Initial Mass Function \citep[IMF,][]{kroupa2001};

We note that some Galactic parameters are still
  very uncertain, in particular many important details remain to be
  settled in the thick disk (scale height, normalization and
  metallicity) and in the halo (normalization) \citep[see e.g.][and
references therein]{Juric2008}. Hence, the comparison plots we are
presenting in the following are
  not meant to provide a best fit to the data, but only a 
  guide to understand the contribution of the various Galactic populations expected in each field.

\subsection{$\omega$~Cen}

\begin{figure*}
\includegraphics[width=0.9\textwidth]{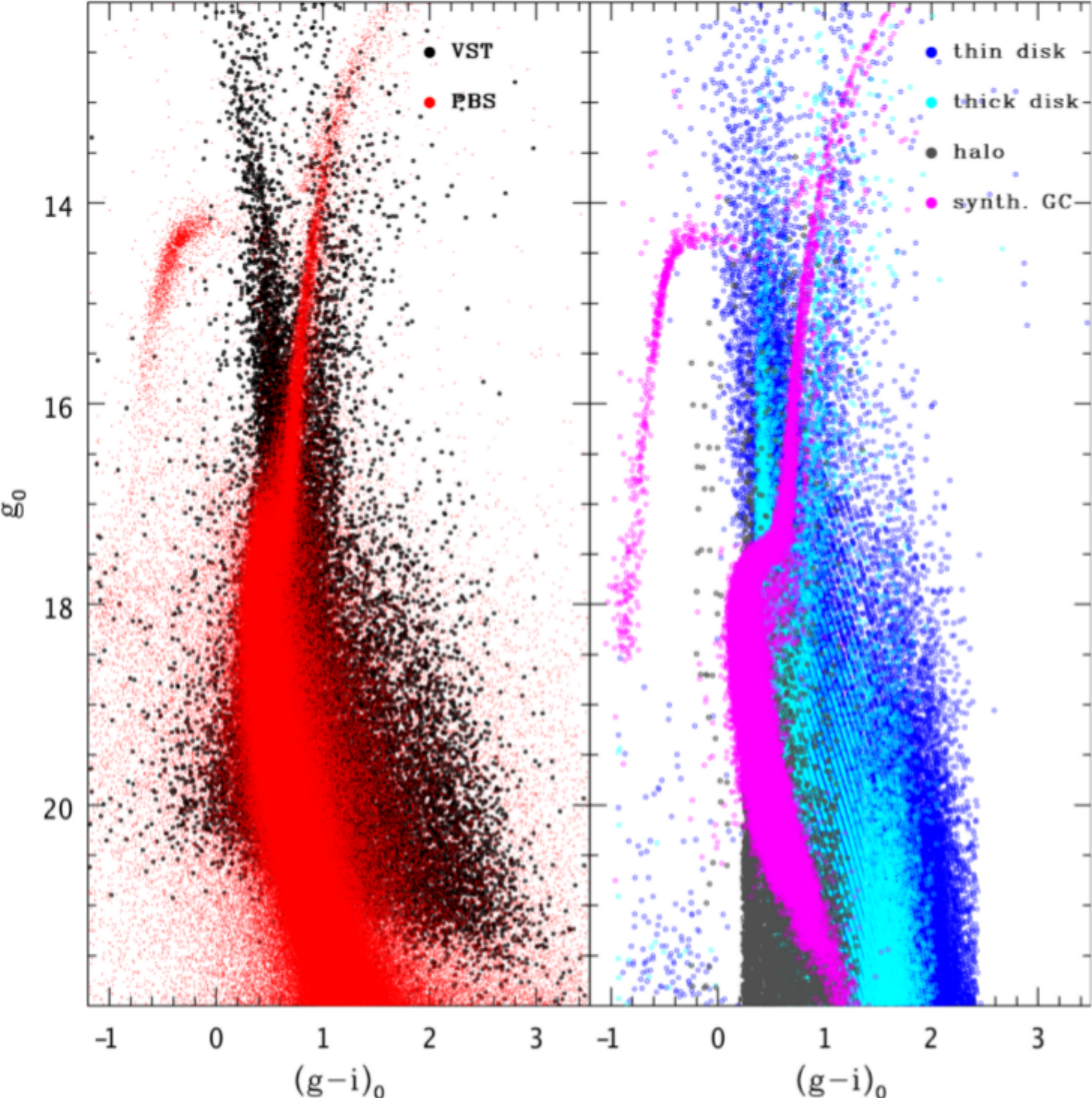}
\caption{Left panel: CMD of stars observed in  Field \#1 around
  $\omega$~Cen (black dots) compared with
the transformed central field photometry by \citet{Castellani07} (red symbols, see
text for details).
Right panel: the Galactic simulation of the same sky area (blue, cyan and grey
symbols  represent thin disk, thick disk and halo stars, respectively)
discussed in the text compared with the synthetic CMD of $\omega$~Cen 
(magenta symbols). See text for details.}
\label{CMD}
\end{figure*}

In the left panel of Fig.~\ref{CMD}, we present the de-reddened CMD
$g_0,(g-i)_0$ for stars (black dots) in Field \#1 (see Fig
\ref{f:omegacen}) around $\omega$~Cen compared with the independent
photometry for the central region by \citet[][red dots]{Castellani07}.
In the right panel of the figure, we show the results of a Galactic
model simulation over 1 deg$^2$ centered on the same field (blue,
magenta and dark grey symbols represent thin disk, thick disk and halo
stars, respectively) compared with the synthetic cluster CMD (red
symbols) computed for $\omega$~Cen (see below for details).


In the Galactic simulation, the photometric errors are not accounted
for. At each given $g$ magnitude, the ``blue edge'' (the bulk of the bluest
stars) can be interpreted as the envelope of the halo MSTO shifted towards fainter magnitudes and redder colors by the
increasing distance and corresponding extinction. At the bright end
($g_0<16$ mag), the CMD is dominated by thin and thick disk stars. 
In particular, in this magnitude range, the blue plume (BP; $0<(g-i)_0<0.75$ mag) is mainly composed
of thin (and a few thick) disk MS stars, whereas the red plume (RP;
$1.2<(g-i)_0<2.0$ mag) is composed of red giant branch stars of both 
disks. At fainter magnitudes, the number of thick disk and halo
stars increases considerably, until it dominates the star counts for
magnitudes fainter than $g_0\approx 18$ mag.

To assist in the interpretation of the observations, we also computed the synthetic CMD of $\omega$~Cen from
the main-sequence phase up to the asymptotic giant branch phase using the
stellar population synthesis code SPoT \citep[Stellar Population
Tools, see for details and ingredients][and references
therein]{brocato2000,raimondo2005,raimondo2009}.  The synthetic CMDs were randomly
populated using Monte Carlo methods with the IMF of \citet{kroupa2001} in the mass interval 0.1--10
$M_{\sun}$. For each star the luminosity and effective temperature
were derived according to evolutionary tracks taken from the BaSTI
database, then magnitudes in the ACS/HST (the Advanced Camera for
  Surveys aboard the Hubble Space Telescope) and Sloan photometric systems
were computed, using color-temperature relations derived from the
stellar model atmospheres by \citet[][and references
therein]{castelli03}. In order to reproduce the multipopulations in
$\omega$~Cen, we used the subpopulation parameters and
ratios suggested by \citet{joolee13} as a first guess. Then,  we considered the ACS/HST
photometry by Sarajedini et al. 2007\footnote{The Globular Cluster
Treasury program (PI: Ata Sarajedini, University of Florida) is an
imaging survey of Galactic globular clusters using the ACS/WFPC
instrument on board the Hubble Space Telescope. Photometric data are
publicly available at http://www.astro.ufl.edu/ata/public\_hstgc}, and tested on the
HST-CMD the {bf Horizontal Branch (HB)} star distribution as derived from the adopted
population ratios. We also explored different assumptions on the
parameterizations of the HB star distribution \citep[the Reimers' mass-loss
parameter $\eta_{\rm R}$, its dispersion $\sigma_{\rm R}$, etc;][]{reimers77}. From
this procedure we derived our best-fit synthetic CMD for the HST data,
which is constructed by assuming the metallicities and ages for six
subpopulations, namely: [Z=0.0003, Y=0.245, t=13 Gyr], [Z=0.0006,
Y=0.245, t=13 Gyr], [Z=0.0008, Y=0.400, t=12 Gyr], [Z=0.0032
Y=0.400, t=12 Gyr], [Z=0.0065, Y=0.400, t=11.5 Gyr],  [Z=0.0165,
Y=0.400, t=11 Gyr] \citep[e.g.][and references therein]{joolee13}. The population ratios are 40\%,
27\%, 15\%, 10\%, 7\% and 2\%, respectively. Note that the adopted metallicity
values and abundance ratios agree with the most recent spectroscopic
observations of RGB stars in the cluster
\citep[e.g.][]{jp10,marino11a,calamida09}.
Each simulation also took into account the photometric error of the adopted HST ACS photometry. We used the
same set of parameters to compute the GC CMD in the Sloan filters,
also shown
in the right panel of Fig.~\ref{CMD}.  It is worth noting that,
according to these simulations, the bluest MSTO might be distinguished
from the Galactic disk population (see also discussion in
Sect. \ref{preliminary}).

\subsection{NGC6752}


\begin{figure*}
\includegraphics[width=0.9\textwidth]{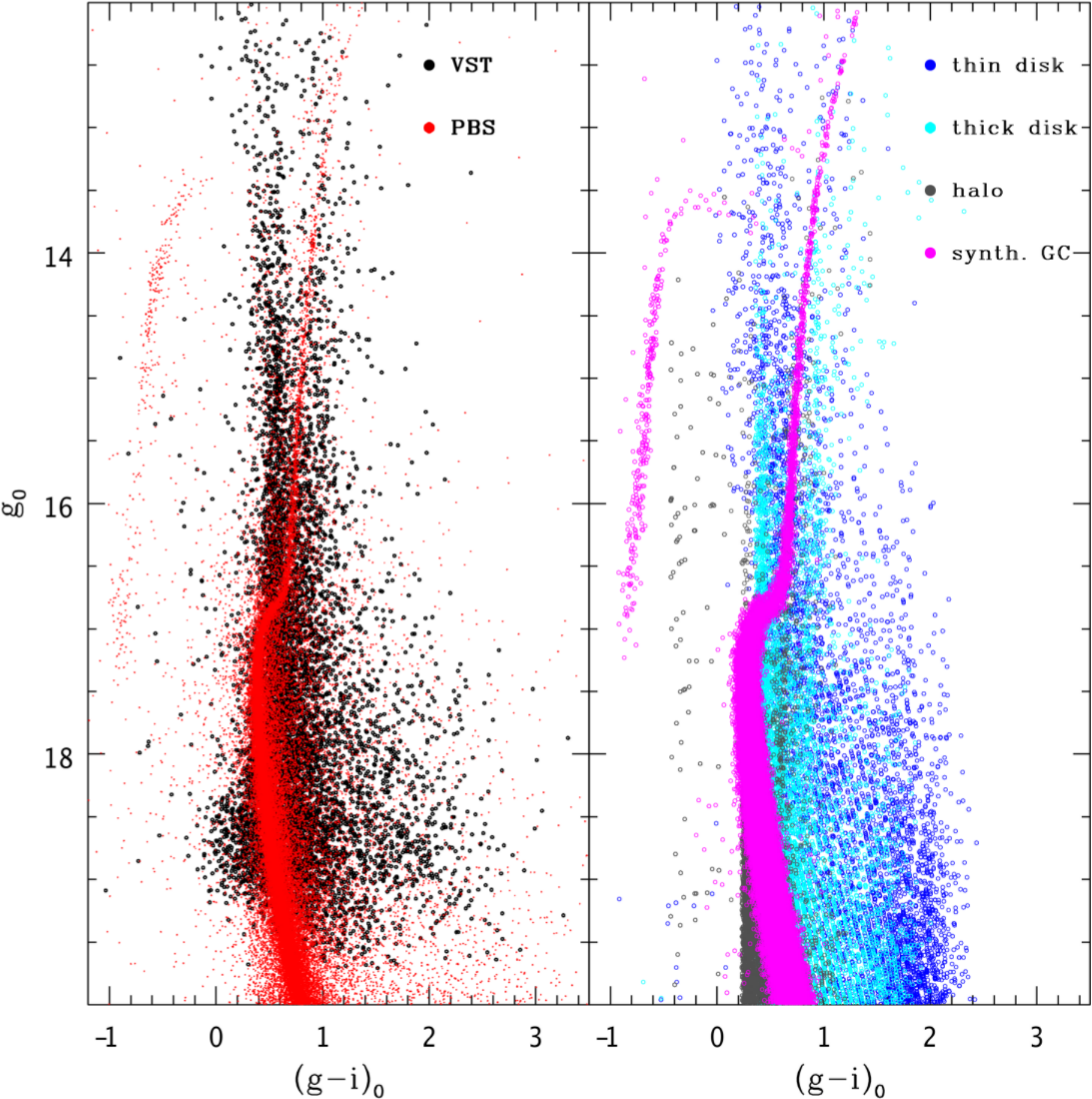}
\caption{The same of Fig.~\ref{CMD} but for the Field \#1 around
  NGC6752. The photometry of the central area of the cluster is based
  on  PBS unpublished data.}
\label{CMDconsimulngc6752}
\end{figure*}

In the left panel of Fig.~\ref{CMDconsimulngc6752},  we present the
de-reddened CMD  $g_0,(g-i)_0$ for stars (black dots) in
Field \#1 (see Fig \ref{f:ngc6752}) around NGC6752 compared with the
independent photometry for the central region by PBS (red dots;
for details, see the previous section).  In the right panel of the figure, the
Galactic simulation for the selected field (same colors as in Fig.~\ref{CMD}) is compared with the synthetic CMD, again generated using the
SpOT code). This synthetic CMD  fits
the CMD and the number of HB stars from ACS/HST observations by
\citet{sarajedini2007}, considering a single-burst stellar population
with $Z = 0.0006$, including  the $\alpha$ element enhancement (in agreement with the spectroscopic metallicity
determination by \citet{gratton05} , namely $[Fe/H]=-1.48$ dex and
$[\alpha/Fe]=$0.27 dex). We used the same set of
parameters to predict the CMD in the Sloan filters, shown in the right
panel of Fig.~\ref{CMDconsimulngc6752}. We note that in order to fit
the very  blue HB of NGC6752
\citep{buonanno1986} we  fixed the chemical composition and  adopted a
high value of the Reimers' parameter $\eta_R=0.68$, increasing the mean
mass loss along the RGB to $\sim 0.28 M_\odot$. Alternatively, in the
emerging multipopulation
scenario for GCs,  these stars could be explained assuming they belong
to a second generation of stars, with different chemistry
\citep[see][and references therein]{Milone2013}.


\begin{figure*}
\includegraphics[width=0.9\textwidth]{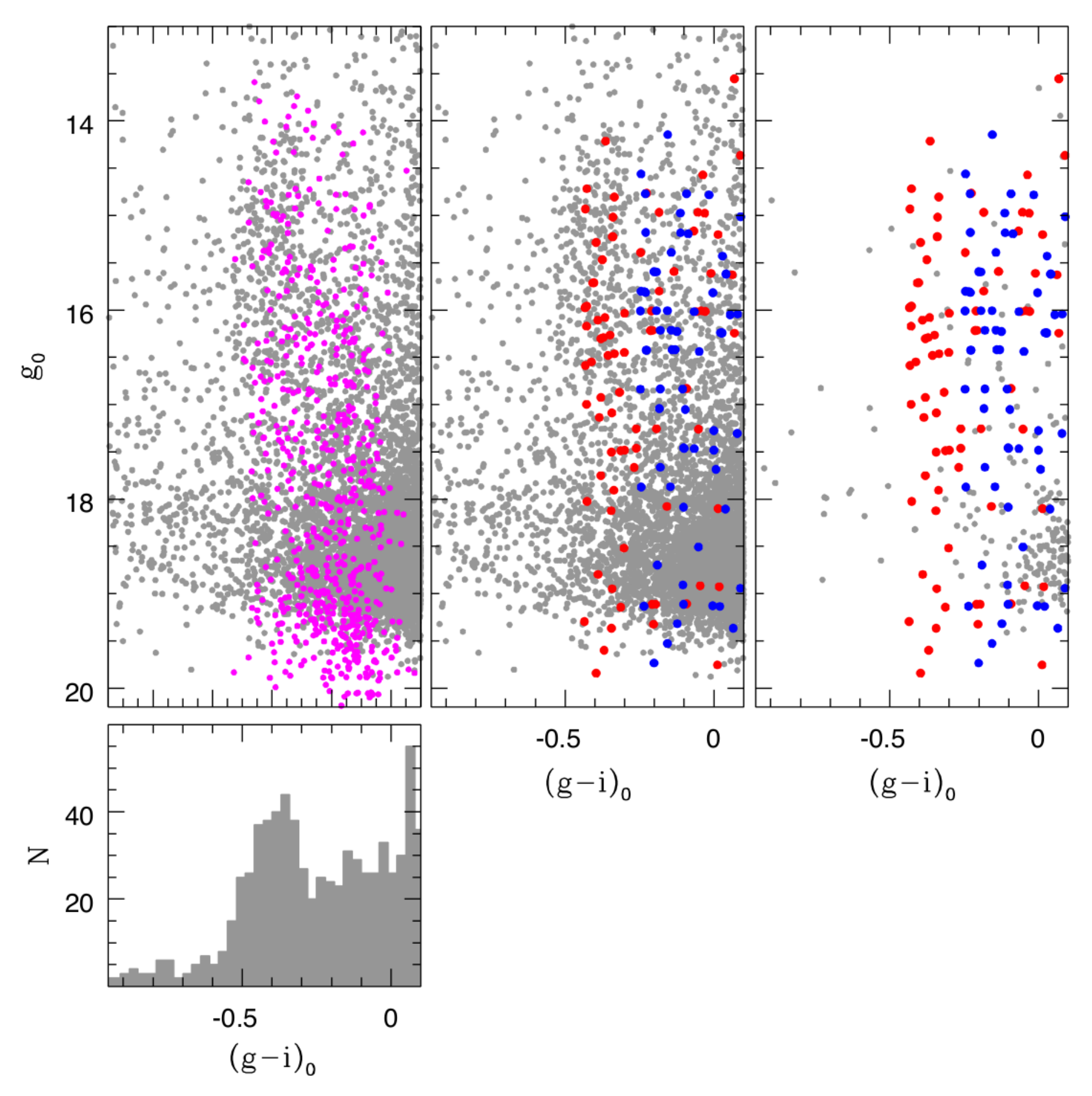}
\caption{A zoom of the cumulative (left and middle upper panels) and  single
  field (right-upper panel) CMD  around NGC6752. The lower panel shows the corresponding
  $g-i$ color histogram for the cumulative CMD. The magenta dots, in
  the left panel,
  represent the SDSS Halo HB stars (see text for details). Red and
  blue dots, in the middle and right panels, are simulated
  Halo HB stars adopting different assumptions on the mass loss (see text).}
\label{sinteticoNGC6752}
\end{figure*}

In the CMD of  Fig.~\ref{CMD_tuttoNGC6752} we can also
identify a vertical structure corresponding to $-0.5<(g-i)_0<-0.2$ mag and
$g_0>14$ mag (see box). The stellar nature of all the objects within this
structure has been visually verified. Their position in the CMD
suggests that they belong to the HB phase associated either with NGC6752 
or with the Galactic halo.    The comparisons  in
Fig.~\ref{CMDconsimulngc6752} show that the observed and predicted
colors of the cluster HB stars are much bluer than the average  color
of the observed vertical structure that is instead reproduced by the
Galactic halo simulation.  A zoom of this portion of the CMD (box in
Fig.~\ref{CMD_tuttoNGC6752}) is shown in the upper panels of 
Fig.~\ref{sinteticoNGC6752}  for the cumulative case (left and middle
panels) and
Field \#1 (right panel), compared with the predicted halo HB stars in
this single field.
In the lower panel of
Fig.~\ref{sinteticoNGC6752}, we plot the $(g-i)_0$ color histogram of this
region for the cumulative CMD, showing a clear peak in the color range of the identified
structure. 
In the left upper panel, as an observational check, we compare this feature with the
location of the halo HB stars identified by the SDSS \citep[magenta
dots;][]{Smith2010}. 
The good agreement, despite the different areas covered by the two surveys, confirms
our hypothesis that these stars belong to the Galactic halo.
As a theoretical check, in the middle and right upper panels, we overlaid two halo simulations
obtained adopting an RGB mass loss with $\eta_R=$ 0.2 (blue dots) and 0.3
(red dots). As expected, the former value produces redder HB
stars, as a consequence of the higher average stellar mass populating
the synthetic HB. We note that the simulation with $\eta_R=0.3$ is more
in agreement with  the observed vertical structure than the one with $\eta_R=0.2$. This discrepancy could be
reduced adopting a metallicity value lower than $Z=0.0002$.
Concerning the predicted number of HB stars, our models outnumber the
observed star counts by roughly a factor two (see
Fig.~\ref{sinteticoNGC6752}). We tentatively ascribe this difference
to the assumed halo/thin disk normalization \citep[1/500;
e.g.][]{bahcall84,Tyson88}, which is probably too high. Indeed, this
number is strongly debated and different values have been proposed,
i.e., 1/1250 \citep[see e.g. ][]{cohen1995}, 1/850 \citep[see
e.g., ][]{Min1998,Morrison1993} or 1/200 \citep[see
e.g., ][]{Juric2008}.  As a bottom line, we stress how the location and
the number of these observed halo HB stars can help to constrain the
physical and numerical assumptions adopted in Galactic simulations.
In particular, the observed number in our sample seems to suggest a significantly
smaller halo/thin disk normalization, close to the lowest values in the
literature \citep{cohen1995,Min1998,Morrison1993}.

\section{Star counts around $\omega$-cen}\label{preliminary}

In this section, we present our analysis of the star counts in the 37
fields covering $\omega$~Cen from the cluster center to about three
tidal radii. 

\begin{figure*}
\includegraphics[width=0.9\columnwidth]{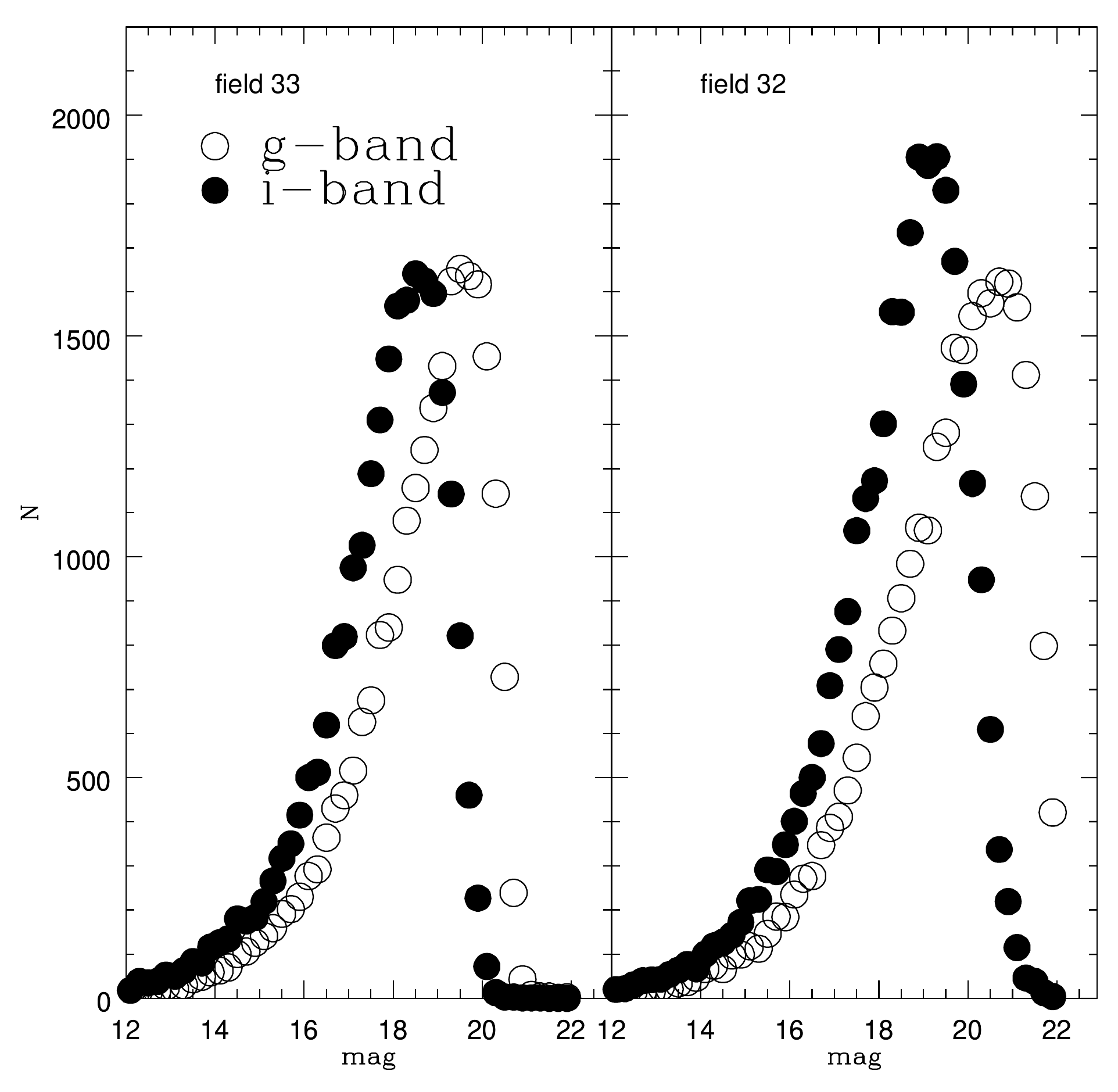}
\caption{Luminosity function of stars in Field \#33 (left panel) and \#32
(right panel) around $\omega$~Cen with a magnitude bin of  0.2 mag.} \label{completeness}
\end{figure*}

The first step to obtain reliable star counts is an evaluation of the
photometry completeness in the adopted bands.
To this purpose, we adopt the same
method used by \citet{Col04,coleman2005}. A luminosity function
was generated for each field  in both the $g$ and $i$ bands adopting a
magnitude bin of 0.2~mag. We then
assumed the completeness limit of each field to be 0.2~mag brighter
than the turnover in the luminosity function.  The overall magnitude
limit of the survey, for each group of fields, is then defined by the
field with the shallowest depth. For $\omega$~Cen we obtain the limit
magnitudes of $g_0 = 19.4$ mag and $i_0 = 18.4$ mag from the luminosity
function of the shallowest field (shown in the left panel of
Fig.~\ref{completeness}).   In the right panel of the figure,  we
 plot the luminosity functions obtained from one of the many fields observed
in good seeing conditions (as an example field \#32), to conclude that
these fields observed in good weather conditions  can be considered complete
down to $g_0=20.8$ mag and $i_0=19.4$ mag.


\begin{figure*}
\includegraphics[width=0.9\textwidth]{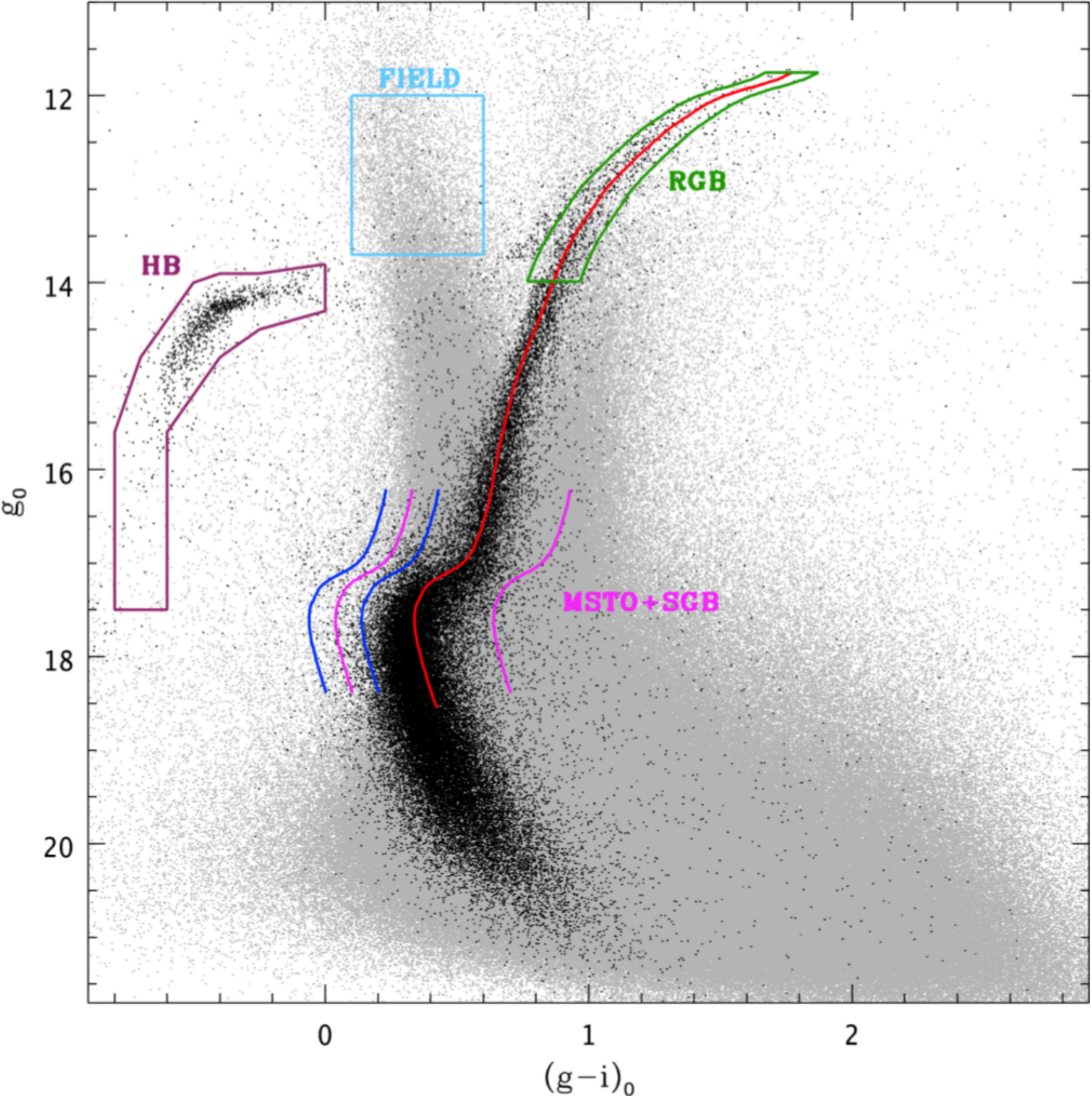}
\caption{Central CMD (black dots) for $\omega$~Cen compared with the cumulative one (grey
  dots). The over-imposed red line is the empirical ridgeline. Stellar evolutionary phases and a FIELD region are labelled (see text for
details).}
\label{cmd_sel}
\end{figure*}

A method to search for stellar over-densities is
deriving star counts in specific CMD regions. The availability of the
central fields, allowed us to use the 1 square degree CMD centered on
$\omega$~Cen  (hereinafter  \rm{``central CMD''}) to define the expected location of the HB, MSTO and RGB
phases. In Fig.~\ref{cmd_sel}, we show the \rm{central CMD} (black
dots) compared with the total CMD of Fig.~\ref{CMDtuttoOMEGA} (grey dots).  
The red line represents the empirical ridge-line obtained from
the observed \rm{central CMD}.
After this procedure, we conservatively 
selected as possible members of the cluster the sources lying within
a defined color range  around the ridge line, that is: 
\begin{itemize}
\item $\pm 0.1$ mag, with
$g_0<14$ mag and $(g-i)_0>0.7$ mag, for the RGB (region inside the green
contour line), 
\item $\pm 0.3$ mag (between the blue lines, to account for
the dispersion of the MSTO region and for  possible photometric
misalignment due to residual differential reddening and/or photometric
errors), with
$16.2<g_0<18.5$ mag\footnote{The fainter end is based on the limiting
magnitude derived from the completeness test discussed above}, for the
MSTO and the Sub Giant Branch (labelled in the following as MSTO+SGB). 
\end{itemize}
For the HB
phase, due to its specific morphology, different ranges in magnitude
and color were considered in order to properly cover the
whole extent (region inside the purple contour line). 
In this CMD, we also identified  a region without cluster contamination
(region inside light blue contour line labelled as FIELD),
corresponding to $12.0<g_0<13.7$ mag and $0.1<(g-i)_0<0.6$ mag, to quantify
the contribution due to the Galactic thin and thick disk (see right
panel of Fig.~\ref{CMD}).


\begin{figure*}
\includegraphics[width=0.9\textwidth]{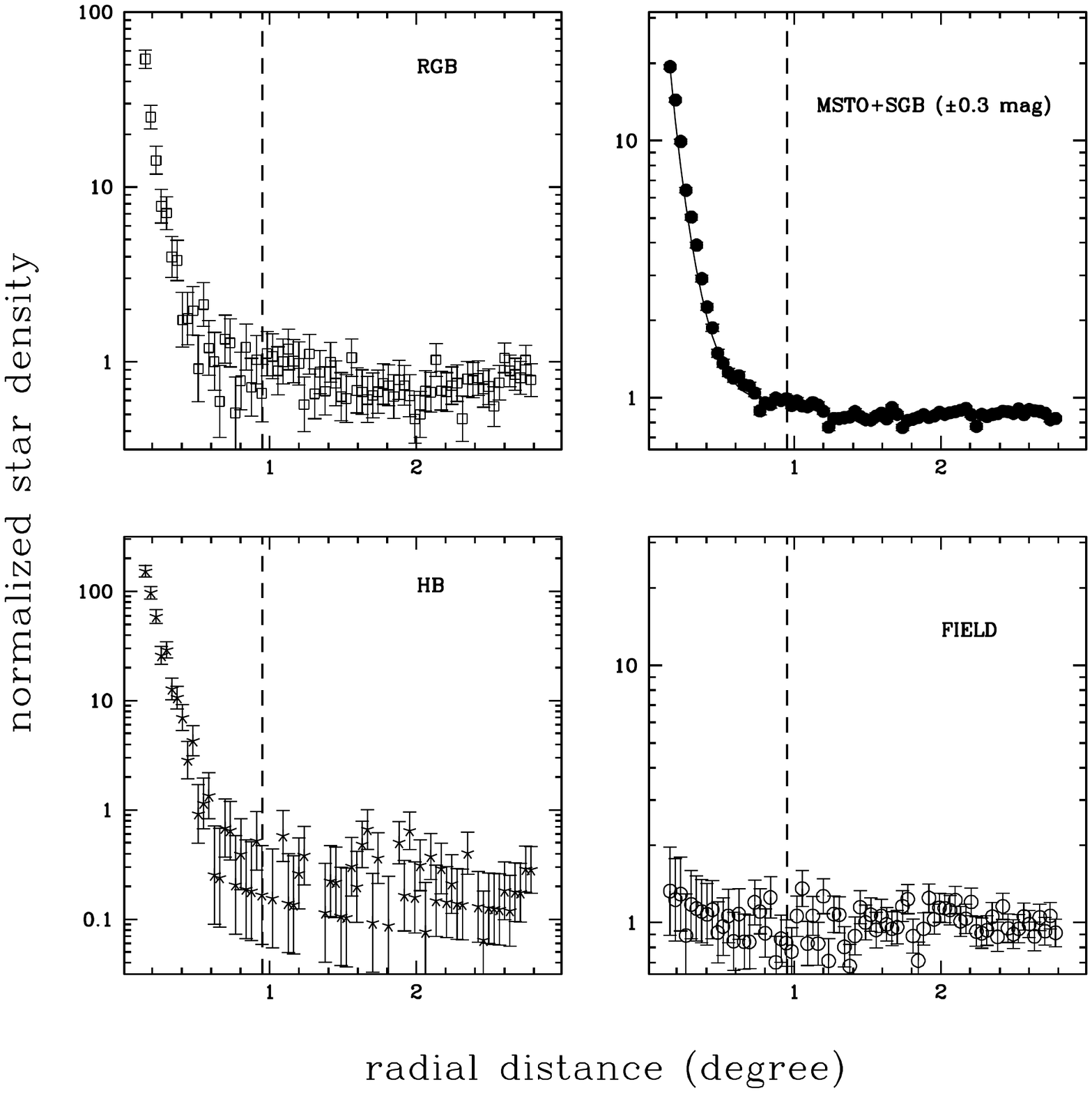}
\caption{Radial profiles of the normalized (see text for details)
star densities, in logarithmic scale, for $\omega$~Cen (selected from the CMD $g$ vs $g-i$), in the three labelled evolutionary phases (HB, RGB,
MSTO+SGB) compared with
the results in the FIELD region. The dashed line represents the nominal
tidal radius, whereas the solid line, in the upper right panel,
  represents the King profile based on Harris (1996) values.}
\label{contcirc}
\end{figure*}


\begin{figure*}
\includegraphics[width=0.9\textwidth]{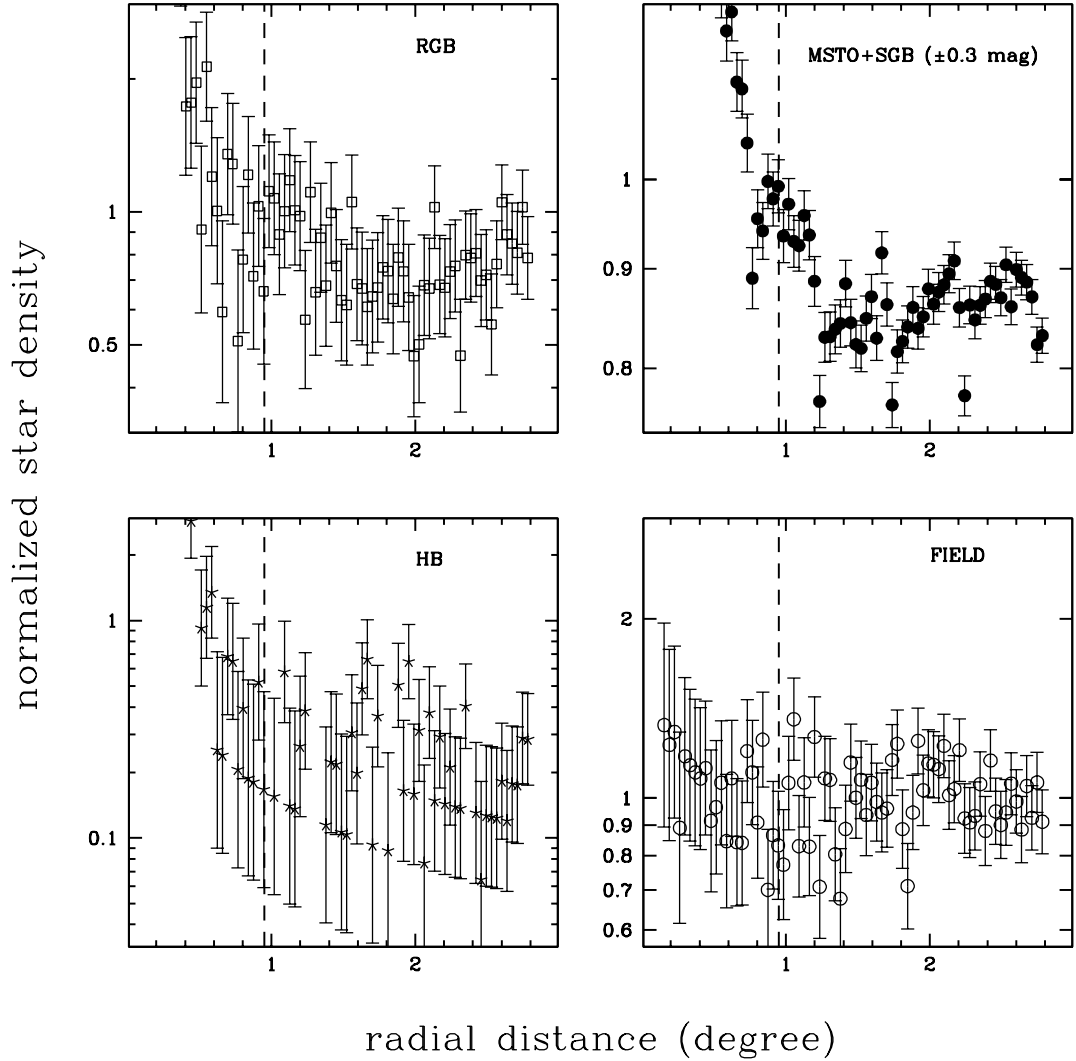}
\caption{A zoom-in of Fig.\ref{contcirc} in the region around the
  nominal tidal radius.}
\label{contcirczoom}
\end{figure*}


\begin{figure*}
\includegraphics[width=0.9\columnwidth]{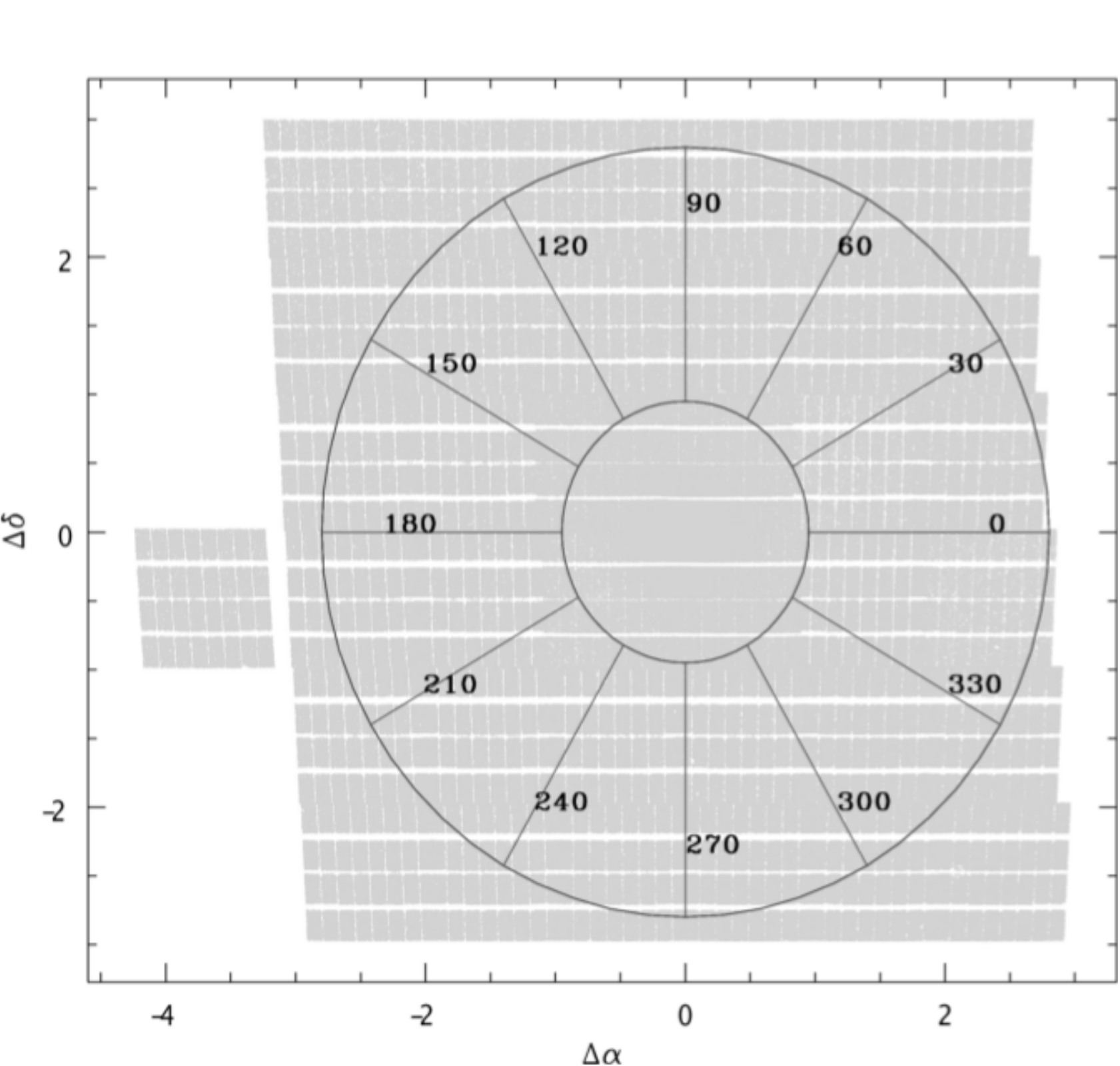}
\caption{Schematic view of the circular sectors used for angular star counts.}
\label{angle}
\end{figure*}

\begin{figure*}
\includegraphics[width=0.9\columnwidth]{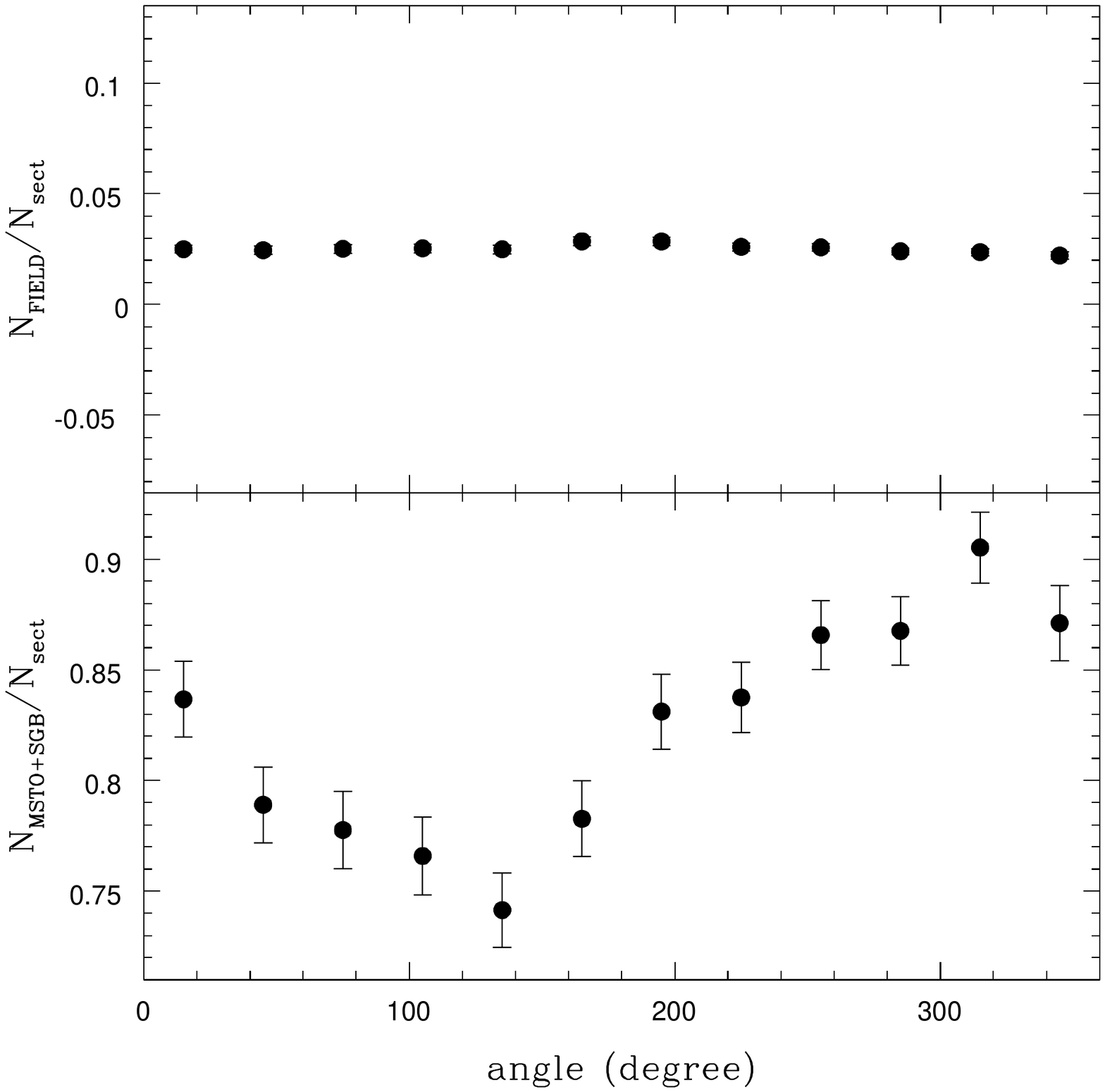}
\caption{Normalized (see text for detail) star counts for $\omega$~Cen (selected from the CMD $g$ vs $g-i$)
  in circular
  sectors for stars in the MSTO+SGB phase within $\pm 0.3$ mag from
  the ridgeline (lower panel) and in the
  FIELD region (upper panel).}
\label{contsetc_0.3}
\end{figure*}

\begin{figure*}
\includegraphics[width=0.9\columnwidth]{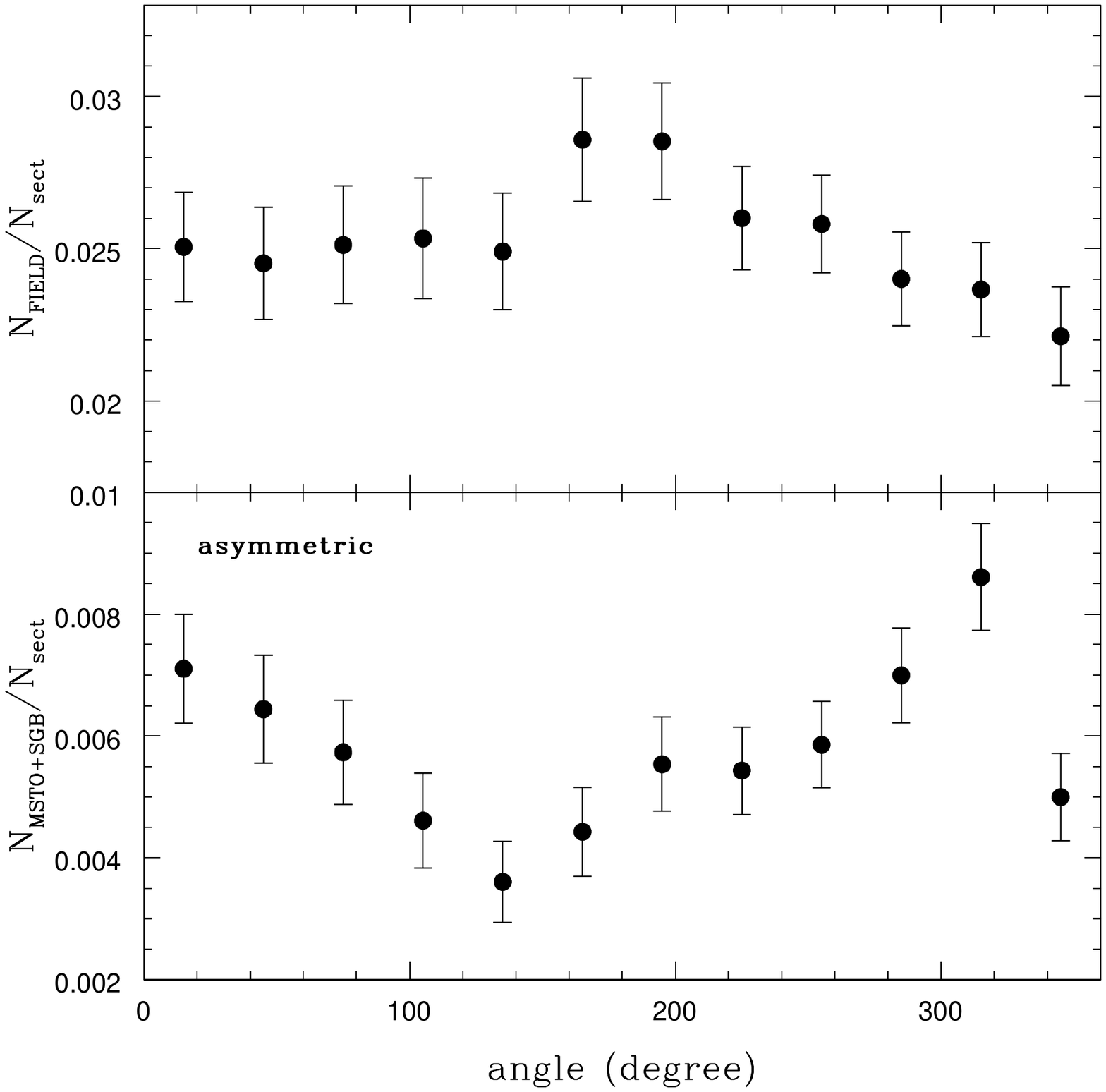}
\caption{Normalized (see text for detail) star counts for $\omega$~Cen (selected from the CMD $g$ vs $g-i$)
  in circular
 sectors for stars in the MSTO+SGB phase, bluer than the ridgeline by
  0.3 to 0.4 mag and fainter than 17 mag (lower panel, labelled as ``asymmetric'') and in the
  FIELD region (upper panel).}
\label{contsetc_0.2_0.4_17}
\end{figure*}

In the panels of Fig.~\ref{contcirc} we plot the radial profiles of the normalized
star densities in the  three evolutionary phases mentioned (HB, RGB,
MSTO+SGB) compared to those in the FIELD region where we expect a
negligible contribution of cluster
stars. We adopted 75 equally spaced annuli, from
the center up to 2.8 deg, corresponding to about 3 tidal
radii. For each selected evolutionary phase, star density in
the $i$-th annulus is normalized to the corresponding 
total value computed in the circle with the radius of 2.8 deg.
The vertical dashed line represents the nominal tidal radius of 0.95
deg \citep{Harris96,dacoco08}, that is intermediate between the two values
provided by \citet{mcl05} adopting a King model \citep[0.80 deg;][]{king66} and a
Wilson model \citep[1.2 deg;][]{wilson75}. We remind that the
measurement of the tidal radius is significantly model-dependent
\citep{mcl05} and difficult to obtain on the basis of simple
structural models due to the possible presence of tidal tails or other
kinds of extra-tidal stellar population \citep[see e.g.][and
references therein]{dice13}.  
In Fig.~\ref{contcirc}, the error bars result from the
propagation of the Poisson errors on the individual counts. A zoom
of these plots to
emphasize the behavior around the tidal radius is shown in
Fig.~\ref{contcirczoom}. As expected,
the FIELD stars do not show any specific trend, whereas in the other
three cases, we obtain the typical expected profile for a globular
cluster \citep[e.g.][see discussion below]{king66, wilson75}. This is shown in the right upper panel of Fig. ~\ref{contcirc}  where the King profile based on \citet{Harris96} values (solid line) is overimposed on the MSTO+SGB star counts.
However, an inspection of Fig.~\ref{contcirczoom} suggests that, apart from the
poorly populated HB phase, the nominal tidal
radius seems to be slightly underestimated with an excess of stars at
this distance from the center, suggesting a better
agreement with the value based on the Wilson model \citep[1.2 deg][]{mcl05}. Moreover, 
we notice a non-negligible increase of the density 
around and beyond 2 tidal radii, both for RGB
and MSTO+SGB stars, thus not excluding the presence of
extra-tidal cluster stars. However, the RGB counts are affected by
large errors due to the smaller statistics, and for this reason, in the following, we concentrate on
the much more numerous MSTO+SGB stars. 
To further investigate the possible presence of tidal tails, we also
considered the MSTO+SGB star counts as a function of the direction. To
this purpose, we divided the total explored area into twelve 30 deg
circular sectors  (see Fig.~\ref{angle})  and counted the MSTO+SGB  stars
within each sector with a radial distance ranging from 1 to 3
tidal radii.  The results are shown in the lower panel of
Fig.~\ref{contsetc_0.3}, where star counts are normalized to the total
number of stars in the same area.  This normalization allows us to
reduce spurious effects related to the differential contribution of
the disk population that is expected to be more important at lower
galactic latitudes (see discussion in section \ref{sec-results}). 
In Fig.~\ref{contsetc_0.2_0.4_17}, we present the same plot, but considering a
different selection of the MSTO+SGB stars, labelled as asymmetric,
that includes stars bluer than
the ridgeline by 0.2 to 0.4 mag (region between the blue lines in Fig.~\ref{cmd_sel}) with
magnitude fainter than 17 mag. This
selection has the drawback of containing a lower number of stars, but
the advantage of reducing the disk contamination, instead including the
cluster MSTO stars emerging from the disk population as shown in
Fig.~\ref{cmd_sel}.  In the upper panel of Figs. \ref{contsetc_0.3}
and \ref{contsetc_0.2_0.4_17}, we show for comparison the
behavior of the same  normalized counts for the FIELD stars defined
above, adopting the same vertical scale used for the
lower panel.
We emphasize that in both figures, the MSTO+SGB star counts show a clear peak around
300 deg (the south-east direction), corresponding to
the predicted ellipticity orientation of the cluster \citep[see e.g. Table
5 in][]{andvan10}. Even if the asymmetric selection is less statistically
significant, it is interesting to note that the trend is the same and
that this effect cannot be related to a larger disk contribution in
this direction because FIELD stars show an opposite behavior.


\begin{figure*}
\includegraphics[width=0.9\columnwidth]{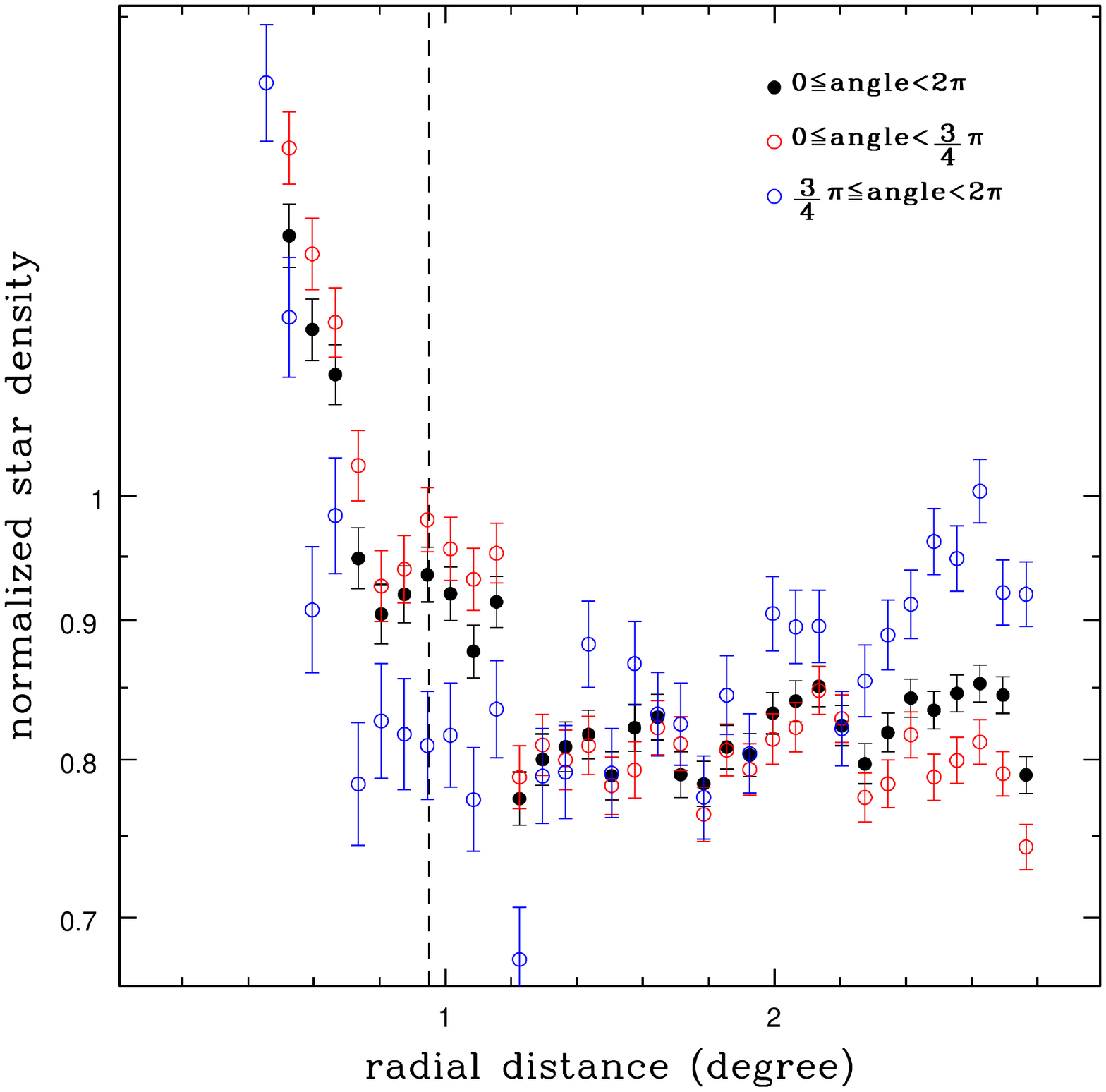}
\caption{Radial profiles of the normalized star densities, in
  logarithmic scale, for the
  $\omega$~Cen 
  MSTO+SGB stars in the labelled angular regions.}
\label{contcirc_conf_sub}
\end{figure*}

The contribution of this direction to the detected extra-tidal
overdensity is also evident from Fig.~\ref{contcirc_conf_sub}
where we  compare the obtained radial profile for MSTO+SGB stars
(black filled circles) with the ones obtained considering the
overdensity direction (blue open circles) and its complement to $2\pi$
(red open circles), respectively. We note that: i) the radial profile
varies with the selected angle with the overdensity detected beyond
the tidal radius mainly due to stars concentrated in the south-east
direction; ii) even excluding the over-density direction, the radial
profile (red open circles) indicates a tidal radius of about 1.2 deg,
with an excess of stars at about 1 deg from the center.


\begin{figure*}
\includegraphics[width=0.9\textwidth]{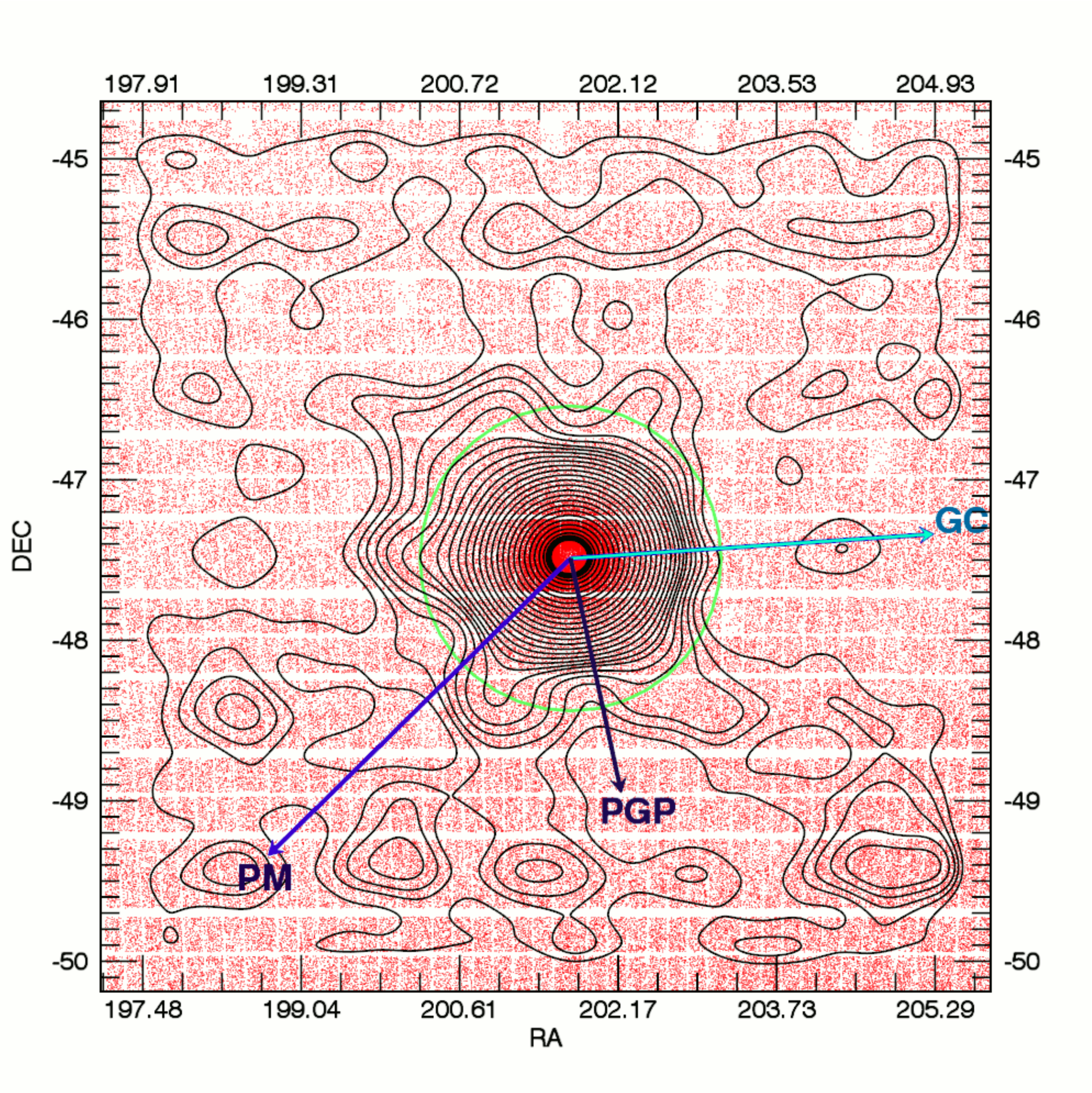}
\caption{ 
Contour levels of candidate $\omega$~Cen stars projected onto the sky. 
The coordinates and the orientation 
are the same as in Fig.\ref{f:omegacen}. The red dots display the radial distribution 
of the entire photometric catalog. The black contours show different 
density level after the subtraction of a plane. The green circle shows 
the truncation radius ($r_t$=0.95 deg). The light blue arrow points to
the Galactic Center (GC), the purple arrow points to the direction of the 
proper motion of the cluster (PM), while the blue arrow the direction 
perpendicular to the Galactic Plane (PGP). Note the clear excess of
star across and 
beyond the truncation radius and the over-densities located in the fourth 
and in the second quadrants. See text for more details.}  
\label{iso_map}
\end {figure*}

To further constrain the nature of the candidate cluster stars located 
beyond the truncation radius we decided to adopt the contour levels. 
Fig.~\ref{iso_map} shows the count levels for the MSTO+SGB ($\pm 0.3$
mag) candidate cluster 
stars (red dots). In order to overcome the steady increase in the 
counts of field stars when moving from the northern to the southern 
direction we performed a simultaneous fit of a surface and a plane. 
We divided the sky area covered by the data into a grid (800$\times$800) 
and subtracted point-by-point the plane from the surface. The contour 
levels plotted in the above figure display several interesting features. 


{\em i)} -- The countour levels display a well defined excess of star 
counts beyond the truncation radius (green circle, $r_t$=0.95
deg). This evidence is clearer in the II, III and IV 
quadrants where the excess extends up to two truncation radii.  
This finding soundly confirms the evidence of extra-tidal stars based 
on the radial density profile.\par  

{\em ii)} -- The contour levels become less and less symmetric 
when moving from the innermost to the outermost cluster regions, in
good agreement with the eccentricity of $\omega$~Cen 
provided by \citet{bianchini13}. \par    

{\em iii)} -- The countour levels display a clear over-density  
when moving from the bottom right to the top left, even if 
the over-density is less clear in the II than in the IV quadrant.   
It is worth mentioning that these overdensities appear to be orthogonal 
to the direction of the proper motion of the cluster \citep[PM, purple arrow;][]{vanleu,vandeven06} and located between the 
direction of the Galactic Center (GC, light blue arrow) and the projection 
on the sky of the direction perpendicular to the Galactic Plane (PGP,
black arrow). \par  

{\em iv)} -- The countour levels indicate that the position angle of
the major axis measured east from north 
is $\sim 140$ degrees. The current estimate agrees quite well with the determination 
based on ACS photometry of the innermost cluster regions provided by \citet{andvan10}. \par   

The above empirical evidence indicates that candidate cluster stars have 
a complex distribution at radial distances larger than two tidal
radii. 
Firmer conclusions concerning the nature of the extra-tidal stars  
are hampered by the lack of information concerning their proper 
motion and radial velocity.


\begin{figure*}
\includegraphics[width=0.9\textwidth]{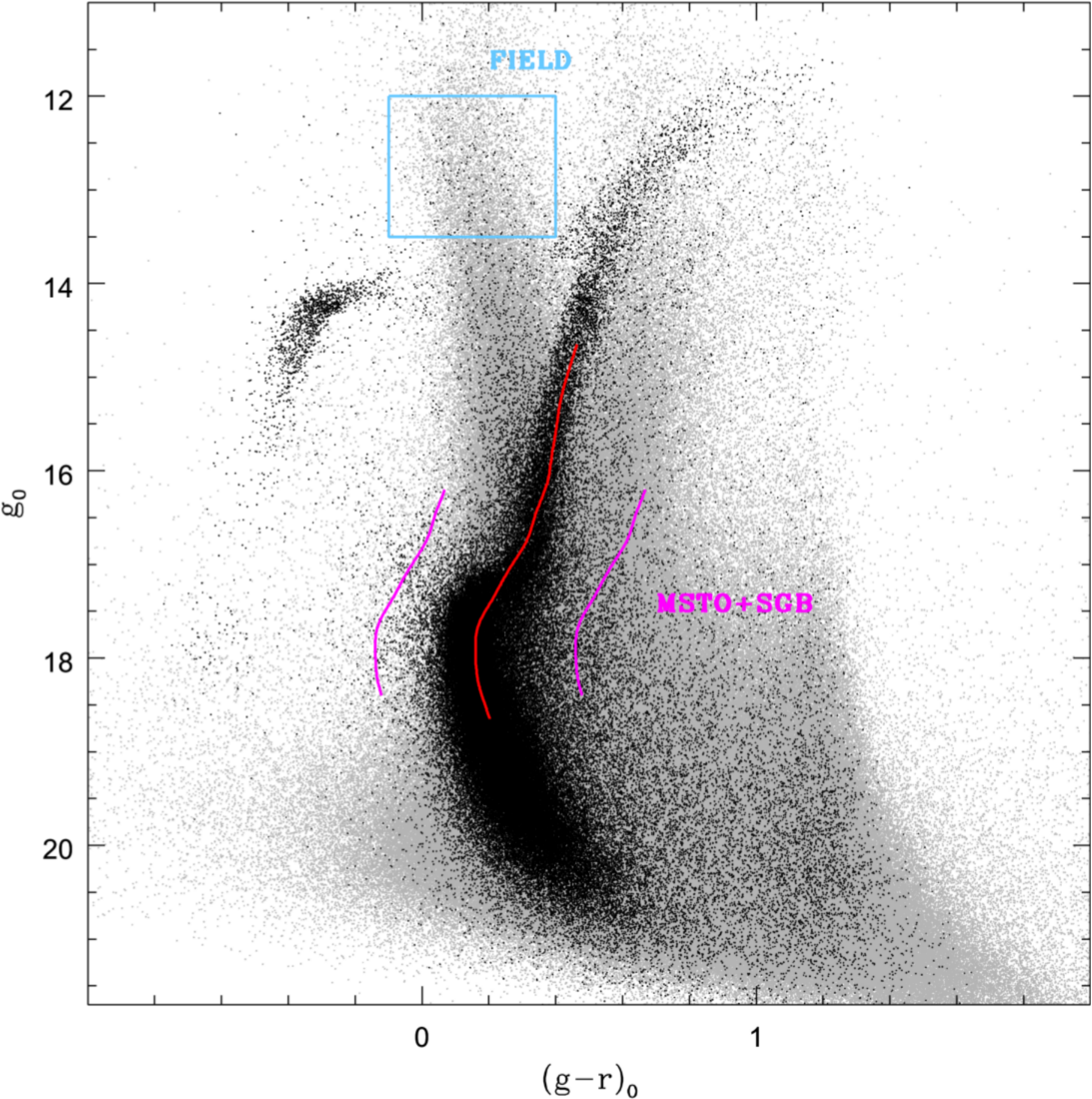}
\caption{Central CMD in $g$ vs $g-r$ (black dots) for $\omega$~Cen compared with the cumulative one (grey
  dots). The over-imposed red line is the empirical ridgeline. The
  MSTO+SGB phase (within the two magenta lines) and a FIELD region
  (cyan box) are labelled (see text for
details).}
\label{cmd_gr_sel}
\end{figure*}


\begin{figure*}
\includegraphics[width=0.9\textwidth]{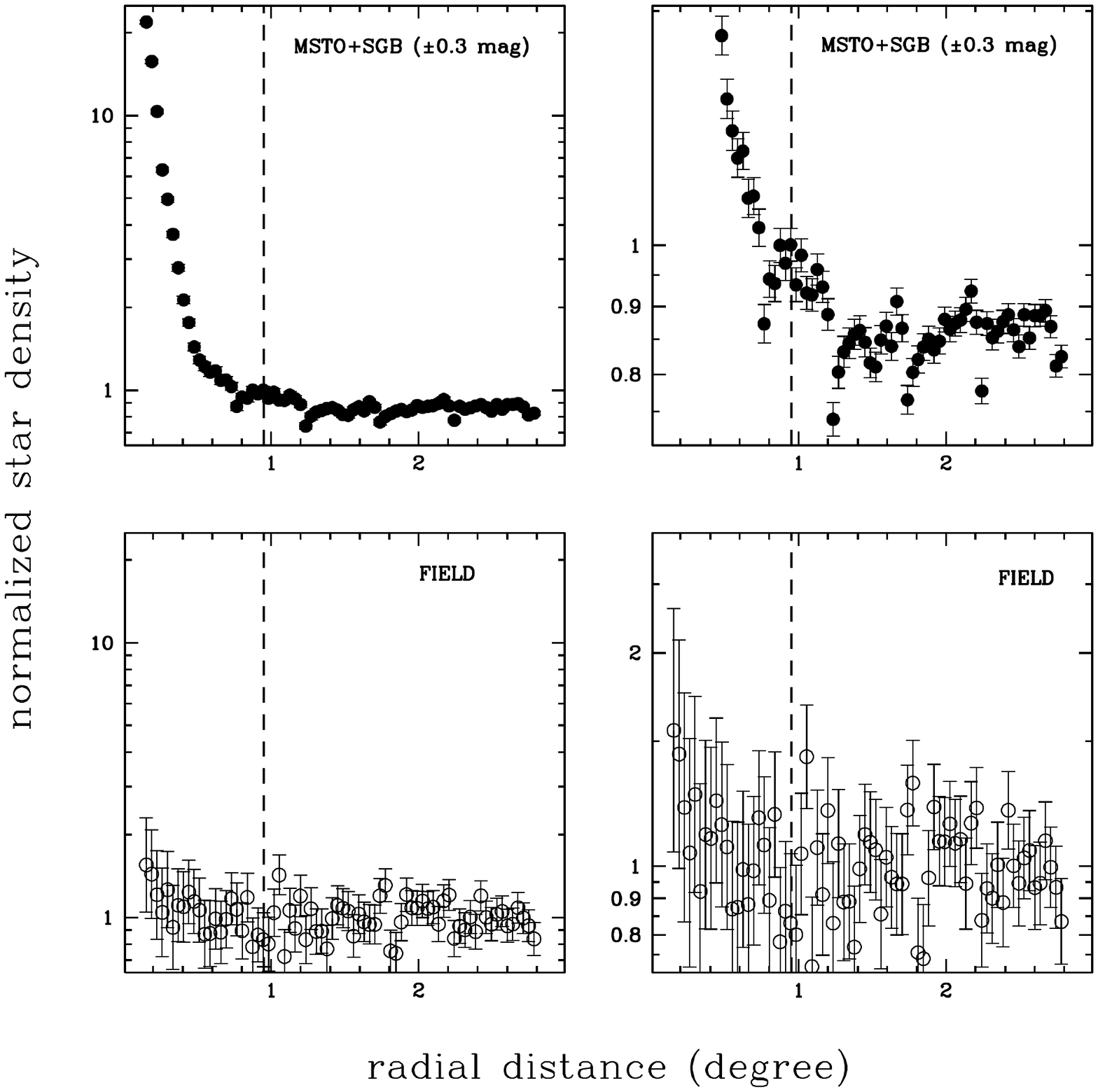}
\caption{Left panels: radial profiles of the normalized (see text for details)
star densities, in logarithmic scale, for $\omega$~Cen (selected from the CMD $g$ vs $g-r$) in the MSTO+SGB evolutionary phase compared with
the results in the FIELD region. Right panels: a zoom-in of the left
panels around the nominal tidal radius (dashed line).}
\label{contcirczoom_gr}
\end{figure*}

\begin{figure*}
\includegraphics[width=0.9\columnwidth]{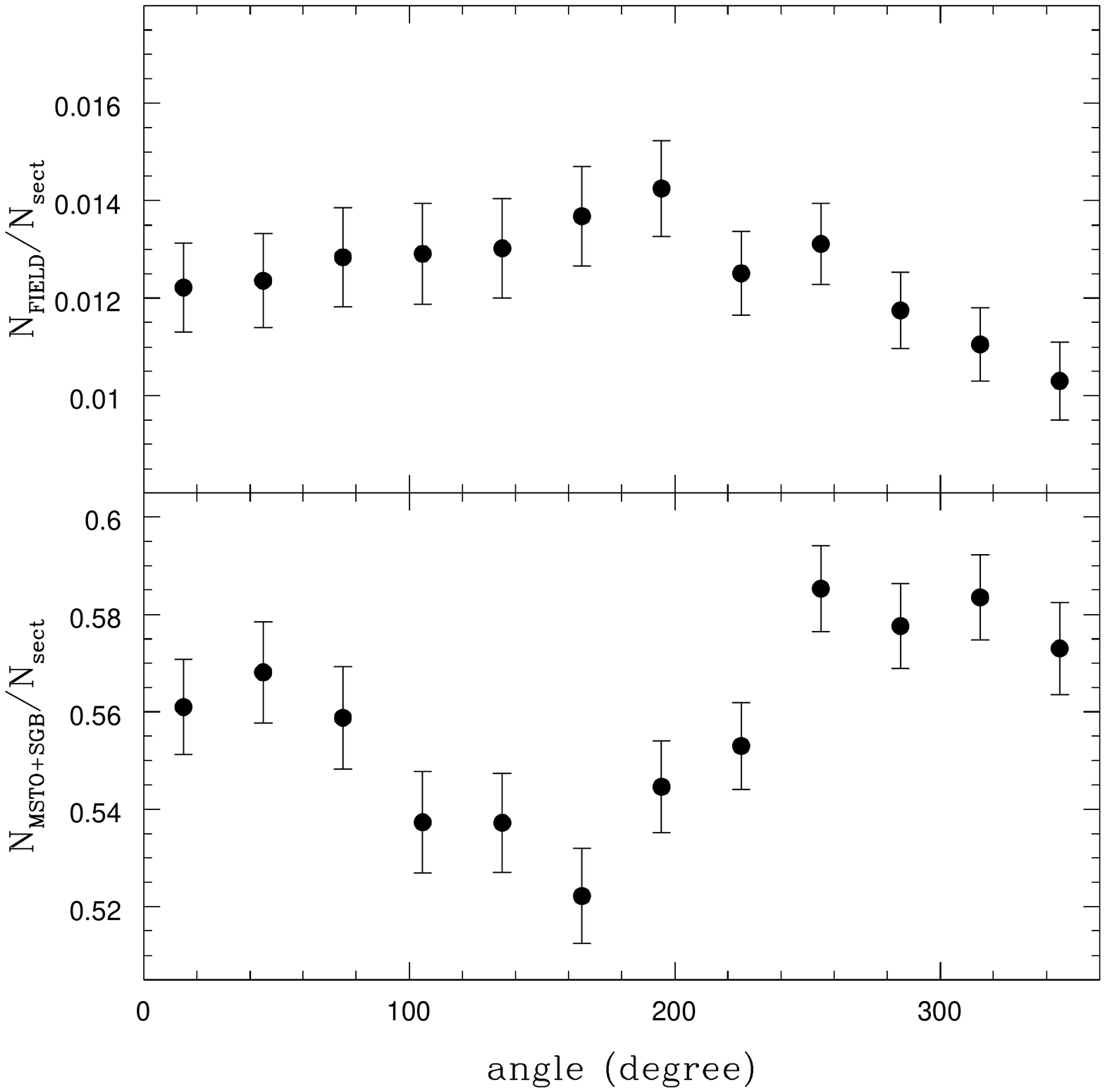}
\caption{Normalized star counts for $\omega$~Cen (selected from the CMD in $g$ vs $g-r$),  in circular
  sectors for stars in the MSTO+SGB phase within $\pm 0.3$ mag from
  the ridgeline (lower panel) and in the
  FIELD region (upper panel).}
\label{fig_contsetc_0.3_gr}
\end{figure*}

As a check of the results obtained in the $g, i$ filters, we perform a star-count
analysis also in the $g$ vs $g-r$ CMD, but only in the more populous MSTO+SGB region. Figure \ref{cmd_gr_sel} shows the
CMD obtained in these filters and the selected regions for star counts
around the cluster MSTO+SGB ($\pm$ 0.3 mag, magenta lines) and in the
FIELD (light blue box). 
These star counts are used to build
the radial profile (see Fig.~\ref{contcirczoom_gr}) and the angular distribution  
(see Fig.~\ref{fig_contsetc_0.3_gr}) in analogy with the
procedure adopted in the $g$ and $i$ filters.
The left panels of Fig.~\ref{contcirczoom_gr} show the
normalized star density as a function of the radial distance from the
cluster center in the MSTO+SGB region (upper panel) and in the FIELD
zone (lower panel). The right panels show the corresponding zoom
plots.
These results confirm the evidence, discussed above, that the cluster tidal radius is closer to the
estimate based on the Wilson model than to the nominal value
(dashed line). 
The angular distribution shown in Fig.~\ref{fig_contsetc_0.3_gr}
supports the evidence of an overdensity in the same directions
identified in Fig.~\ref{contsetc_0.3}.

                                
These results seem to suggest the presence of a tidal tail in addition
to a general extension of the cluster beyond the traditionally assumed
tidal radius. 
We notice that the presence of extra tidal stars support previous evidence by \citet{Meza05} concerning a peak of angular momentum, consistent with that of $\omega$~Cen, identified in the stellar samples by \citet{Beers00} and \citet{Gratton03}.

A similar
behavior was also claimed by \citet{Maj12} with the direction of the
$\omega$~Cen debris mentioned in their paper consistent with
 the one identified in this paper \citep[see Fig.~1a
in][]{Maj12}. Obviously, these preliminary results need additional
investigation and spectroscopic confirmation.

\section{Summary and Future Perspectives}\label{sec-conclusions}

We have presented an overview of the STREGA survey on the VST
Guaranteed Time, focused on a multifilter photometric investigation
of selected regions of the Galactic halo.
The core program of the survey is focused on the investigation of
neighboring areas of a number of GGCs and dSphs reaching at least 2
tidal radii in different directions with the final aim of detecting
extratidal stellar populations either in tails or in extended haloes.
The adopted stellar tracers of streams and haloes are both MSTO and
variable stars but so far only runs involving the former have
been  completed, and data reduction and analysis performed, for three GGCs, namely  $\omega$~Cen, NGC6752
and Pal~12. 
For the first two systems we have also performed a comparison with
model predictions and discussed some preliminary
results. The most relevant ones are:
\begin{itemize}
\item  A possible structure of  Galactic halo HB stars is present in
  the CMD around NGC6752, barely consistent with the presence of
  cluster HB stars, and their number and location can be used to
  constrain the assumptions of Galactic models. The observation of the
  four central square degrees of NGC6752 in the next ESO Period of
  observation will allow us to produce the radial profile of star
  counts in the CMD and to investigate in detail the possible
  presence of tidal tails and/or of an extended halo around this system;
\item We confirm that the nominal tidal radius for $\omega$~Cen  seems to be
  underestimated, suggesting a value more in agreement with 1.2 deg as
  based on the Wilson model. 
\item  We find evidence of stellar over-density around  $\omega$~Cen at about 1 deg from the
  center in the North-West direction and at about 2 deg in the
  opposite direction. The resulting asymmetric, elongated extra-tidal
  structure seems to be consistent with the extension to longer radial
  distances of current measurements and
  predictions for the cluster ellipticity profile and orientation.
\end{itemize}

In the case of Pal~12 the interpretation of the CMD and the star count
investigation is in progress (Di~Criscienzo et al. in prep).
In the next two periods of observations (ESO P93 and P94) the
time-series data of
the selected fields for Pal~3, Fornax, and Sculptor should be
completed and we will be able to use RR~Lyrae stars in combination
with MSTO stars to identify extratidal stellar populations around
these systems, up to at least three tidal radii.
The narrow band observations for the WDs and IBs in the Pal 3 fields
will also be performed.
In the subsequent two semesters we plan to extend the analysis to Phoenix and
Sextans, thus completing the core program of the Survey.

\section{Acknowledgments}
We thank our anonymous referee for her/his valuable comments and suggestions.
This work was supported by PRIN–INAF 2011 ``Tracing the formation and
evolution of the Galactic halo with VST'' (PI: M. Marconi), PRIN-INAF
2011 ``Galaxy Evolution with the VLT Surveys Telescope (VST)'' (PI: A. Grado) and 
PRIN–MIUR (2010LY5N2T) ``Chemical and dynamical evolution of the Milky
Way and Local Group galaxies'' (PI: F. Matteucci). MDC acknowledges the
support of INAF through the 2011 postdoctoral fellowship grant on the STREGA survey.

{}


\end{document}